\documentclass[jcp,amsmath,amssymb,reprint,twocolumn]{revtex4}

\usepackage[dvips]{graphicx}
\usepackage[dvips]{color}
\usepackage{subfigure}
\usepackage{epsfig}       
\usepackage{dcolumn}      
\usepackage{bm}           
\usepackage{amsmath} 
\usepackage{amssymb} 
\usepackage{longtable}
\usepackage{enumitem}
\usepackage{multirow} 

\begin{document}

\title{Breakdown of the Stokes-Einstein relation in two, three and four dimensions}
\author{Shiladitya Sengupta$^{1,2}$, Smarajit Karmakar$^{2}$, Chandan Dasgupta$^{3}$, Srikanth Sastry$^{1,2}$}
\affiliation{
$^{1}$ Theoretical Sciences Unit, Jawaharlal Nehru Centre for Advanced Scientific Research, 
Jakkur Campus, Bangalore 560 064, India.\\ 
$^{2}$ TIFR Centre for Interdisciplinary Sciences, 21 Brundavan Colony,
Narsingi,Hyderabad 500075, India.\\
$^{3}$ Centre for Condensed Matter Theory, Department of Physics, Indian Institute of
  Science, Bangalore, 560012, India.}
\date{\today}

\begin{abstract}

The breakdown of the Stokes-Einstein (SE) relation between
diffusivity and viscosity at low temperatures is considered to be one
of the hallmarks of glassy dynamics in liquids. Theoretical analyses
relate this breakdown with the presence of heterogeneous dynamics, and
by extension, with the fragility of glass formers.  We perform an
investigation of the breakdown of the SE relation in 2, 3 and 4
dimensions, in order to understand these interrelations. Results from
simulations of model glass formers show that the degree of the
breakdown of the SE relation decreases with increasing spatial
dimensionality. The breakdown itself can be rationalized {\it via} the
difference between the activation free energies for diffusivity and
viscosity (or relaxation times) in the Adam-Gibbs relation in three
and four dimensions.  The behavior in two dimensions also can be
understood in terms of a generalized Adam-Gibbs relation that is
observed in previous work.  We calculate various measures of
heterogeneity of dynamics and find that the degree of the SE breakdown
and measures of heterogeneity of dynamics are generally well
correlated but with some exceptions.  The two dimensional systems we
study show deviations from the pattern of behavior of the three and
four dimensional systems both at high and low temperatures.  The
fragility of the studied liquids is found to increase with spatial
dimensionality, contrary to the expectation based on the association
of fragility with heterogeneous dynamics.
\end{abstract}

\pacs{xx}

\maketitle

\section{Introduction}
The dramatic slowdown of dynamics upon cooling glass-forming liquids
towards the glass transition is described by the temperature
dependence of a number of transport coefficients, and relaxation time
scales, which include the shear viscosity ($\eta$), the translational
diffusion coefficient ($D$), the rotational correlation time
($\tau_{c}$) and the structural relaxation time $\tau_{\alpha}$
obtained from the long time decay of density correlation
functions. The Stokes-Einstein (SE) relation \cite{book:SE-HM,
  pap:Einstein, book:LL-FM} in its original formulation relates the
translational diffusion coefficient ($D$) of a macroscopic, or
Brownian, probe particle - a single particle property - to the shear
viscosity ($\eta$) of the liquid - a collective property at a
temperature $T$ : $D = \frac{mk_{B}}{c \pi R}\frac{T}{\eta}$ where $m$
is the mass and $R$ is the radius of the particle, $T$ is the
temperature of the liquid and the factor $c$ ($ = 6$ or $4$) depends
on the (stick or slip) boundary condition at the surface of the
Brownian particle \cite{book:SE-HM}.  Although a hydrodynamic
relation, the SE relation is known to be applicable even for the self
diffusion of the liquid particles at high temperatures
\cite{pap:Hodgdon-Stillinger}. However, several experiments and
simulation studies in the last three decades \cite{pap:SE-Pollack,pap:SE-Fujara-etal, pap:SE-Chang-Silescu, pap:Heuberger-Silescu,pap:SE-Cicerone-Ediger, pap:SE-Blackburn-etal, pap:SE-Mapes-etal,pap:Swallen-etal, pap:Swallen-etal-PRL03, pap:SE-Rossler,thesis:Ashwin, pap:KOB-MCT, pap:SE-Tarjus, pap:SE-Monaco,pap:SE-Ngai-Magill-Plazek, pap:SE-Hansen-etal,pap:SE-Water-liu-etal,pap:SE-Chen, pap:SE-Becker,pap:SE-Fernandez-alonso-etal,pap:SE-Mallamace-etal-Water,pap:SE-Xu-etal-water-SEB,pap:demichele-leporini,pap:BordatJPCM,pap:SE-Affouard2009,pap:SE-Han-Schober} have conclusively shown that
at low temperatures in supercooled liquids, the SE relation breaks
down, such that self diffusion coefficient is much larger than what
one may infer, using the SE relation, from the value of the
viscosity. Although many works claim that the break down of the SE
relation occurs around the mode coupling temperature $T_{c}$
\cite{ediger-review}, more recent results ({\it e.g.}
\cite{pap:BordatJPCM}) suggest that the breakdown occurs in the
vicinity of a much higher temperature, the onset temperature
$T_{onset}$, at which aspects of slow dynamics including stretched
exponential relaxation (see below) begin to be manifested
\cite{pap:AG-Sastry-Deb-Still,pap:Sastry-pcc,pap:Sastry-physA}. There
is a commonly held view that the breakdown of the SE relation is a
manifestation of dynamical heterogeneity (DH), which is characterized
by various indicators we discuss below, such as the dynamical
susceptibility
($\chi_{4}$)\cite{pap:Kirk-Thiru,pap:4pt-CD,pap:Ovlap-Glotzer-etal,pap:Ovlap-Donati-etal,pap:Karmakar-PNAS,pap:Berthier-etal-DH},
the Kohlrausch-William-Watts (KWW) exponent ($\beta_{KWW}$)
quantifying the stretched exponential decay of correlation functions,
and the non-Gaussian parameter $\alpha_2$ that measures the deviation
of particle displacements from a Gaussian form.
We note, however, that there is no agreement on the nature 
and origin of heterogeneity among the theories proposed to explain the SE
breakdown ({\it e.g.}, dynamical facilitation
\cite{pap:SE-Jung-Garrahan-Chandler}, the random first order
transition theory \cite{pap:SE-Xia-Wolynes,pap:Frag-Xia-Wolynes,pap:SE-Lubchenko-Wolynes}, 
the mode-coupling theory \cite{pap:Biroli-Bouchaud}, the shear transformation
zone theory \cite{pap:SE-Langer} and the obstruction model
\cite{pap:SE-Douglas-Leporini}). 
The goal of the present work is to interrogate the manner 
in which the SE breakdown changes with the spatial dimensionality, 
and the correlation of SE breakdown with characteristics of heterogeneous 
dynamics, and fragility. Since the spatial dimensionality enters explicitly 
in some of the theoretical descriptions mentioned above, this study is expected 
to provide valuable information about the validity of these theories. 
As we discuss below, our investigations reveal 
surprises whose rationalization should lead to a better understanding 
of the key features of glassy dynamics.

The rest of this paper is organized as follows: In Sec. II, we provide
a discussion of the SE relation and its relation with dynamical
heterogeneity and fragility, that forms the necessary background for
our work.  In Sec. III we provide details regarding the models studied
and the simulation procedure. Sec. IV contains the results of our
study. Sec. V contains a discussion and conclusions arising from our work.

\section{Stokes-Einstein breakdown, dynamical heterogeneity and fragility} 

One of the earliest theories of the breakdown of the SE relation is due to Hodgdon and
Stillinger \cite{pap:Hodgdon-Stillinger} who envisage the highly
viscous supercooled liquid to be composed of sparse ``fluid-like''
regions of low viscosity in a matrix of a more viscous fluid. Thus,
both $D$ and $\eta$ are space-dependent. By calculating the viscous
drag force on a diffusing particle in the ``fluid-like'' region, they
showed that the local drag force decreases from the Stokes' value, and
hence the local diffusion coefficient increases. If one uses the bulk
viscosity (which is dominated by the more viscous regions) in the SE relation,
one finds a breakdown of this relation. In order to apply their model to realistic
systems and to explain the difference between the behavior of translational 
and rotational diffusion coefficients,
Hodgdon and Stillinger had to impose certain special properties
on their model. However, Tarjus and Kivelson \cite{pap:SE-Tarjus} later
argued that the mere existence of domains with different local properties is a
sufficient condition for the SE breakdown. They considered a liquid in which
there are domains (of unspecified nature) of size $L$ with a size
distribution $\rho(L)$ such that $\rho(L)L^{2}dL$ is the probability
of finding a molecule in a domain of size between $L$ and $L+dL$. They
assumed that the local relaxation time $\tau(L)$ in a domain is size
dependent and the measured $\alpha$ relaxation time ($\tau_{\alpha}$)
is the average of $\tau(L)$:

\begin{eqnarray}
\tau_{L} &\propto& \exp( E(L) / k_{B} T) \nonumber\\
\tau_{\alpha} = \langle \tau_{L} \rangle &\propto& \int_{0}^{\infty}  \rho(L) \exp( E(L) / k_{B} T) L^{2} dL
\label{eqn:SE-TK-tau}
\end{eqnarray}
In their picture, the SE relation is valid inside a domain. However, the translational diffusion involves 
passage through many domains. This is why the average $D$ is different from the prediction of the SE
relation. Assuming that a diffusing particle performs a random walk across domains and $D(L)$ changes abruptly at interfaces, 
and neglecting a term involving the gradient of $D$, one obtains: 
\begin{eqnarray}
D(L) &\propto& T / \eta(L),\quad  \eta(L) \propto  \exp( E(L) / k_{B} T) \nonumber\\
D = \langle D(L) \rangle &\propto& \int_{0}^{\infty}  \rho(L) D(L) L^{2} dL\nonumber\\
&\propto& \int_{0}^{\infty}  \rho(L) \exp( -E(L) / k_{B} T) L^{2} dL
\label{eqn:SE-TK-D}
\end{eqnarray}
Unless $\rho(L)$ is a $\delta$-function, Eqs. \ref{eqn:SE-TK-tau} and \ref{eqn:SE-TK-D} lead to a 
violation of the normal behaviour, $D\tau=constant$. 

This picture of the heterogeneity may also be interpreted in 
terms of the existence of a distribution of local relaxation times. 
Blackburn {\it et al.} \cite{pap:SE-Blackburn-etal} argued that the translational diffusion 
coefficient $D$ and the rotational correlation time $\tau_{c}$ measure different moments of this distribution, 
thus causing the SE breakdown: 
\begin{eqnarray}
D &\propto& \langle \frac{1}{\tau} \rangle, \quad \tau_{c} = \langle \tau \rangle \nonumber\\
D \tau_{c} &\propto& \langle \tau \rangle \langle \frac{1}{\tau} \rangle \nonumber
\end{eqnarray}
\begin{equation}
\langle \tau \rangle \langle \frac{1}{\tau} \rangle =\left\{ \begin{array}{ccl} 1 
\mbox{ for $\delta$ function distribution, normal SE}\\
\!\!\!>1\mbox{ SE breakdown}\end{array}\right.
\label{eqn:SE-tau-invtau}
\end{equation}
La Nave {\it et al.} showed, \cite{pap:LaNave-etal} using the potential energy landscape framework, 
that in the 3D Kob-Andersen model \cite{pap:KA} the product $<D><1/D>$ indeed grows as the temperature decreases. 
Swallen {\it et al.} \cite{pap:Swallen-etal} argued that since $D\tau_{c}$ increases with decreasing temperature, 
Eq. \ref{eqn:SE-tau-invtau} implies that the distribution of relaxation time should be broader at lower $T$. 

The stretching exponent $\beta_{KWW}$ provides a measure of the width of the distribution of relaxation times. 
Let $\rho(\tau)$ denote the distribution of relaxation times $\tau$ where each of the local relaxation function 
is exponential with relaxation time $\tau$. The overall correlation function $\phi(t)$ is empirically 
given by a stretched exponential:

\begin{equation}
\phi(t) = \int_{0}^{\infty} d \tau \rho (\tau) \exp(-t/\tau)  =  \exp \left(- (t / \tau_{KWW})^{\beta_{KWW}} \right)\nonumber
\end{equation}

Using simple mathematical identities, it can be shown \cite{pap:Colmenero-etal} that the $n$-th moment 
$\langle \tau^{n} \rangle$ of the distribution $\rho(\tau)$ is given by

\begin{equation}
\langle \tau^{n} \rangle = \frac{\tau_{KWW}^{n}} {\beta_{KWW}} \frac{\Gamma \left(\frac{n}{\beta_{KWW}} \right)} {\Gamma(n)}
\end{equation}
Using the above formula, it is easy to show that
the relative variance, which provides a measure of the width of the distribution, is given by:
\begin{equation}
\frac{ \langle \tau^{2} \rangle - \langle \tau \rangle ^{2}} { \langle \tau \rangle ^{2} } = \frac{\beta_{KWW}\Gamma \left( \frac{2}{\beta_{KWW}}\right) }{\Gamma \left(\frac{1}{ \beta_{KWW}}\right)} -  \Gamma \left(\frac{1}{ \beta_{KWW}} \right)
\label{eqn:chp3-width-distr}
\end{equation}
This equation implies that the relative variance depends \emph{only} on $\beta_{KWW}$ and 
increases monotonically as $\beta_{KWW}$ decreases.

The above analysis suggests the following interpretation of the SE breakdown. As the temperature decreases, DH 
(of unspecified nature) develops which leads to the existence of a distribution 
of local relaxation times, the width of which increases as $T$ is decreased. This is manifested as 
(i) the lowering of  $\beta_{KWW}$ and (ii) the SE breakdown. Since both the SE breakdown and
the lowering of $\beta_{KWW}$ from 1 are manifestations of DH, 
they should occur simultaneously according to this interpretation.

We should, however, note the following points that argue against this interpretation: (i) the lowering of $\beta_{KWW}$ does not \emph{prove} 
the existence of a distribution of relaxation times \cite{pap:Colmenero-etal} {\it i.e.} liquids 
at low temperatures can be dynamically homogeneous with a single relaxation time but exhibit an inherently 
non-exponential decay of correlation functions and (ii) some experiments on OTP seem to suggest that $\beta_{KWW}$ 
remains \emph{constant} in the relevant low-temperature range (See \cite{pap:Swallen-etal} and references therein). 
In that case the width of the distribution does not change with temperature, and Eq. ~\ref{eqn:SE-tau-invtau} can not explain 
the observation that the degree of the SE breakdown becomes progressively larger as the temperature is decreased. 

The DH can be quantified by more direct indicators. In the present study, we have used 
the following quantities to characterize the DH.
 
\paragraph{The dynamical susceptibility $\chi_{4}$:} The dynamical susceptibility $\chi_{4}(t)$, which is 
the integral of the four-point correlation function $g_{4}(\vec{r},t)$, measures the \emph{fluctuation} in 
the two-point correlation function $q(t)$ (overlap function) \cite{pap:Karmakar-PNAS}. 
Thus $\chi_{4}(t)$ is a direct measure of 
the DH. For supercooled liquids, $\chi_{4}(t)$ shows a peak at a time proportional 
to the $\alpha$ relaxation time \cite{pap:Karmakar-PNAS}. The peak value of the dynamical susceptibility - 
$\chi_{4}^{\mbox{peak}}$ - is a direct measure of the volume of space correlated during structural 
relaxation \cite{pap:Berthier-etal-DH}. Experiments and simulations, typically carried out in three dimensions, 
show that $\chi_{4}^{\mbox{peak}}(T)$ grows monotonically as the temperature $T$ is lowered, which indicates 
that larger regions of space are dynamically correlated at lower temperature {\it i.e.} the DH 
is more prominent at lower temperature. 

\paragraph{The stretching exponent $\beta_{KWW}$:} The stretching exponent $\beta_{KWW}$ is a measure of how non-exponential 
the decay of a time correlation function is. According to the interpretation of the non-exponential decay 
of the density-correlation function as a manifestation of heterogeneous dynamics, a
lower value of $\beta_{KWW}$ implies stronger DH. 

\paragraph{The non-Gaussian parameter $\alpha_2(t)$:} This parameter quantifies the deviation of the distribution of particle  
displacements in time $t$ from the Gaussian form expected for spatially homogeneous dynamics. The non-Gaussian
parameter, calculated from the second and fourth moments of the distribution of the displacements, exhibits a peak at a 
characteristic time that increases as the temperature is decreased. The peak value of $\alpha_2$ provides a characterization of the
spatial heterogeneity of the dynamics. 

\paragraph{Fragility:} Fragility is a material parameter that measures how rapidly the viscosity (or relaxation time) 
of supercooled liquids increases as the temperature decreases. B\"{o}hmer {\it et al.} \cite{fragility_angell} in 
their extensive compilation of available data found that the kinetic fragility $m$ determined from the slope of the Angell 
plot at $T_{g}$ has negative correlation with the stretching exponent $\beta_{KWW}$, {\it i.e.}, smaller $\beta_{KWW}$ 
implies a higher fragility. This conclusion remains the same for both isobaric and isochoric fragilities \cite{pap:frag-beta-Niss}.  
Such a relation has also been obtained theoretically within the framework of random first order transition theory \cite{pap:Frag-Xia-Wolynes}. 
Based on this correlation between fragility and $\beta_{KWW}$, fragility can be considered to be an indicator of DH - 
\emph{more fragile systems are more heterogeneous}.

However, there are evidences against this correlation as well: (i) In experiments on supercooled water confined 
in nano-pores, Chen {\it et al.} \cite{pap:SE-Chen} observed a fragile to strong transition and found fractional SE 
relations in both the regimes. However, the breakdown exponent is closer to 1 ($=0.74$) for fragile water 
than that for strong water ($=0.67$). (ii) Similarly, the breakdown exponent \emph{computed} by Jung {\it et al.} 
\cite{pap:SE-Jung-Garrahan-Chandler} using dynamical facilitation theory was also closer to 1 ($=0.73)$ for a fragile 
glass-former model than for a strong glass-former model ($=0.67$).  (iii) In \cite{pap:PEL-Heuer} it was shown 
that the strong correlation found by B\"{o}hmer {\it et al.} \cite{fragility_angell} between the kinetic fragility 
and $\beta_{KWW}$ becomes much weaker if  subgroups ({\it e.g.} \emph{only} simple or complex molecular glass-formers) are
considered. (iv) Dyre also claimed \cite{pap:Dyre-tenthemes}, on the basis of experiments on simple, organic glass-forming liquids 
that no clear correlation is present between these two quantities. (v) In the simulation of ST2 water, 
Becker {\it et al.} \cite{pap:SE-Becker} observed a breakdown of the SE relation at low $T$ in both strong and fragile regimes, 
with the breakdown exponent nearly the same for both strong and fragile water. (vi) Vasisht and Sastry \cite{pap:vasisht-SEB},
similarly find in simulations of silicon that the SE breakdown becomes apparent in the high temperature, high density liquid, 
and the breakdown exponent  is the same in both the high temperature (fragile) and low temperature (strong) liquids.

We emphasize, however, that the above picture is based on experiments and simulations predominantly in three dimensions and 
does not, \emph{a priori}, tell us what to expect in other spatial dimensions.
In the present study, we aim to understand the inter-relations among DH, 
the SE breakdown and the fragility in other 
spatial dimensions by studying model glass-forming liquids in 2, 3 and 4 dimensions and 
by considering \emph{both} the SE breakdown and direct measures ($\chi_{4}$, $\beta_{KWW}$, $\alpha_2$) of DH and fragility.

We end this section by remarking on the study of SE relation in two
dimensions. Strictly speaking, transport coefficients such as $\eta$ and $D$ are not well-defined in 
infinitely large two-dimensional systems at equilibrium due to the presence of long-time tails 
\cite{pap:2D_Ernst-Hauge-Leeuwen,pap:2D_Dorfman-Cohen} in correlation functions appearing in 
the Green-Kubo formulae for these transport coefficients, and for similar reasons the the use 
of the Stokes relation is questionable \cite{pap:2D_Douglas}. However, these effects are not 
important for the length and time scales considered in our 2D simulations. We find well-defined 
diffusive behavior (mean-square displacement proportional to time) at long times in all our 2D 
simulations, from which the diffusion coefficient $D$ can be obtained without any ambiguity. 
This is consistent with the results of several existing MD simulations of 2D model systems 
\cite{pap:2D_Liu-Goree, pap:2D_liu-Goree-Vaulina,pap:2D_Gravina-Ciccotti-Holian, pap:2D_Hoover-Posch} 
in which the diffusion coefficient and the viscosity have been computed and the validity of the 
SE relation has been examined \cite{pap:2D_liu-Goree-Vaulina}.

\section{Simulation details}


In the present study we perform NVT MD simulations for the following models : (a) the Kob-Andersen binary mixture at the canonical $80:20$ composition \cite{pap:KA} in 4, 3 and 2 dimensions (denoted by 4D KA, 3D KA, 2D KA respectively) at number densities $\rho=1.60$ (4D), $\rho=1.20$ (3D and 2D); (b) the modified Kob-Andersen model (denoted by 2D MKA) at a different composition $65:35$ \cite{pap:Bruning} at the number density $\rho=1.20$; (c) the 50:50 binary mixture of purely repulsive soft spheres with potential $V(r) \sim r^{-10}$ in 3 and 2 dimensions (denoted by 3D R10 and 2D R10 respectively) at a number density $\rho=0.85$ \cite{pap:r10defnSK}. The details of the potentials and units are described in the corresponding references. The integration time step was in the range $dt \in \left [0.001-0.006 \right]$ depending on the temperature. An algorithm due to Brown and Clarke \cite{pap:BC} was used to keep the temperatures constant. 
System sizes were (1) $N=1500$ for 4D KA; (2) $N=1000$ for 3D KA; (3) $N=1000$ for 2D KA; (4) $N=2000$ for 2D MKA; (5) $N=10000$ for 3D R10 and (6) $N=2048$ for 2D R10. Runlengths at each temperature were at least $100 \tau_{\alpha}$ (the $\alpha$ relaxation time defined below).


We have computed the following measures of time scales : 
\begin{enumerate}
\item
Translational diffusion coefficients $D_{A}$ of one type of particles measured from the mean squared displacement (MSD) of that type of particles.
\item
$\alpha$ relaxation times estimated from the time taken to decay to $1/e$ of the initial value of (a) the overlap function $q(t)$ \cite{pap:4pt-CD, pap:Ovlap-Glotzer-etal,pap:Ovlap-Donati-etal,pap:Lacevic,pap:Karmakar-PNAS}, (b) the intermediate scattering function $F(k,t)$, and (c) the self part of the intermediate scattering function $F_{sA}(k,t)$ for one type of particles. The value of $k$ corresponds to the first peak of the partial structure factor $S_{AA} (k)$ of one type of particles.

Our procedure for computing the overlap function is defined in \cite{pap:Frag-Sengupta}. The intermediate scattering function $F(k,t)$ and its self part  $F_{s}(k,t)$ are defined as 
\begin{eqnarray}
F(\vec{k},t)&=&\sum_{i=1}^{N}\sum_{j=1}^{N}  e^{-\imath \vec{k}\cdot (\vec{r}_{i}(t) - \vec{r}_{j}(0) )}\nonumber\\
F_{s}(\vec{k},t)&=&\sum_{i=1}^{N}e^{-\imath \vec{k}\cdot (\vec{r}_{i}(t) - \vec{r}_{i}(0) )}\nonumber\\
\label{eqn:Fkt}
\end{eqnarray}

\item
The shear viscosity ($\eta$) computed from  equilibrium simulations using the Einstein and the Green-Kubo relations. The Green-Kubo relation for the shear viscosity is given by the integral of the auto correlation function of the stress tensor $P_{\alpha\beta}(t)$ : 

\begin{eqnarray}
\eta &=& \frac{V}{k_{B}T}\int_{0}^{\infty} dt <P_{\alpha\beta}(t) P_{\alpha\beta}(0)>\nonumber\\
P_{\alpha\beta}(t) &=& \frac{1}{V}\left( \sum_{i=1}^{N} p_{i\alpha}p_{i\beta}/m + \sum_{i=1}^{N}\sum_{j > i}^{N} r_{ij\alpha}f_{ij\beta} \right)
\label{eqn:chp3-etaGK}
\end{eqnarray}
where $r_{ij}=|\vec{r}_{i} - \vec{r}_{j}|$ and $f_{ij}=-\frac{\partial U(r_{ij})}{\partial r_{ij}}$ and $\alpha, \beta \in (x,y,z)$ denotes Cartesian components.

The shear viscosity can also be computed from the corresponding Einstein relation as the slope of the Helfand moment at long time and in linear regime : 
\begin{equation}
\eta = \frac{1}{Vk_{B}T} \lim_{t\rightarrow \infty} \frac{<(A_{\alpha\beta}(t) - A_{\alpha\beta}(0))^{2}>}{2t}
\label{eqn:chp3-etaEin}
\end{equation}
where $A_{\alpha\beta}(t)$ is the Helfand moment and related to the stress tensor as 
\begin{equation}
\frac{d A_{\alpha\beta}(t)}{dt} = P_{\alpha\beta}(t) V \nonumber
\end{equation}
Since the shear viscosity is a collective property of $N$ particles, numerical accuracy is a big issue when computing shear viscosity using either Eq. \ref{eqn:chp3-etaGK} or Eq. \ref{eqn:chp3-etaEin}. To improve numerical accuracy we take average over different components. We have found that values from the two methods mutually agree well and report here the shear viscosity obtained from the Einstein method (denoted by $\eta_{Einstein}$). 

\end{enumerate}

\section{Results}

In this section, we describe our results for the SE breakdown,
fragility and different measures of DH in 2, 3 and 4 dimensions. In
Sec. IV (A) we show diffusivity, viscosity and relaxation time data
that demonstrate the SE breakdown. We further show that in 3 and 4
dimensions, the breakdown exponent can be understood in terms of the
Adam-Gibbs relation for diffusivity and relaxation time. 
In 2 dimensions, the Adam-Gibbs relation is not valid as we have previously
shown \cite{pap:AG-Sengupta}. We show here that nevertheless, the
the SE breakdown in two dimensional systems can be rationalized in terms of a generalized
Adam-Gibbs relation observed to hold in these systems. In Sec. IV (B)
we describe various measures by which heterogeneity can be quantified,
and the correspondence of the degree of heterogeneity with the degree
of the SE breakdown. In Sec. IV (C) we show that fragility increases
with increasing spatial dimensionality, contrary to expectations that
the fragility is correlated with the degree of heterogeneity of
dynamics.  We show however that the variation of fragility with
dimensionality can be understood in thermodynamic terms, through the
evaluation of the configurational entropy and properties of the
distribution of local energy minima or inherent structures \cite{pap:AG-Sastry}.

\subsection{The dimension dependence of the SE breakdown}

\begin{figure}[h!]
\begin{center}
\includegraphics[width=8cm,height=6cm]{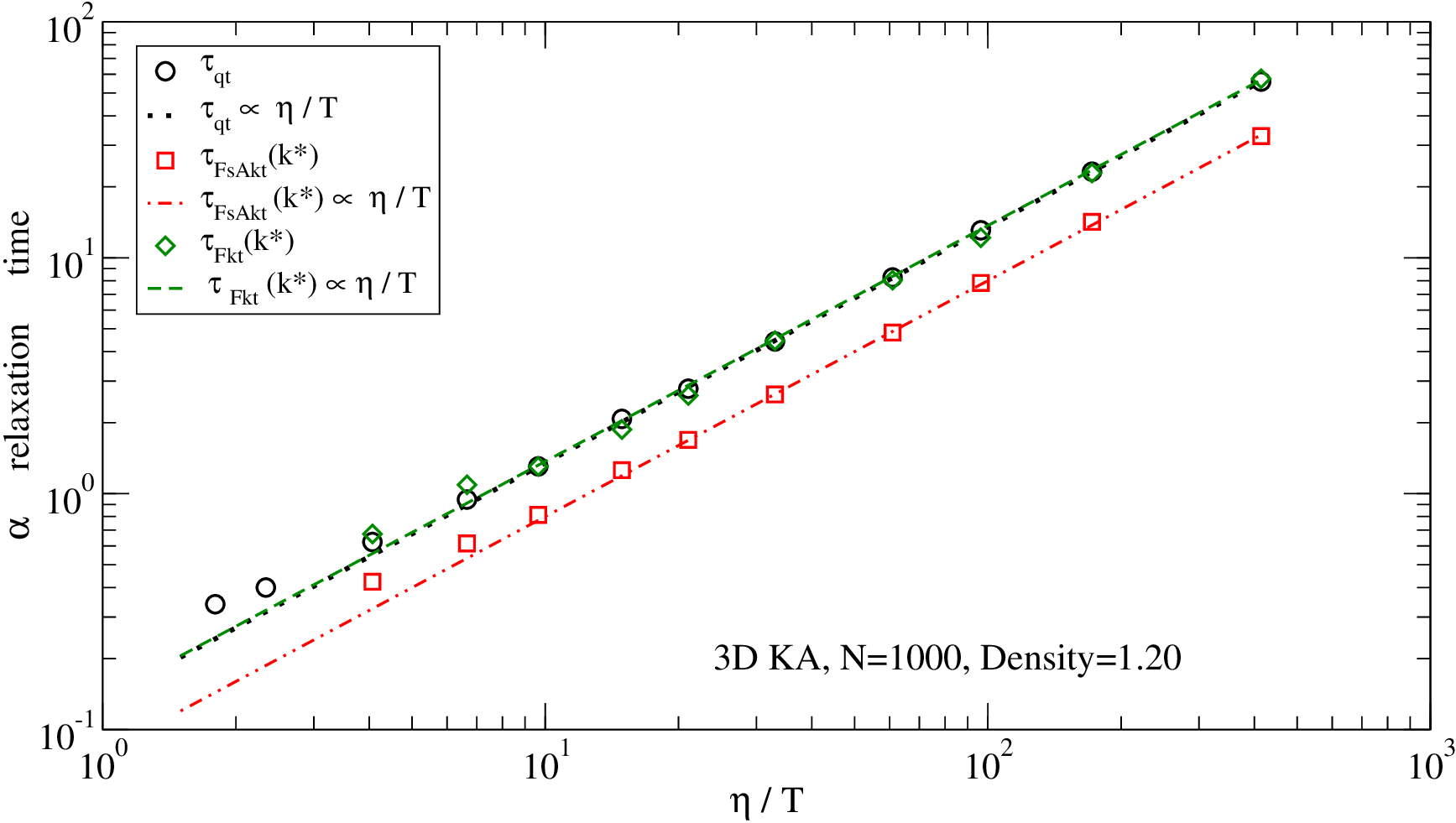}\\
\caption{$\alpha$ relaxation times $\tau$ obtained from (i) the overlap function ($\tau_{qt}$, (ii) the self part ($\tau_{FsAkt}$) of the intermediate scattering function $F_{sA}(k^{*},t)$ of one type ($A$) of particles and (iii) from the full ($\tau_{Fkt}$) intermediate scattering function $F(k^{*},t)$ plotted against viscosity showing that $\tau \propto \eta / T$ is a good description of data at low $T$ in the 3D KA model. Systematic deviations are seen at high $T$. $k^{*}$ is at the first peak of the partial structure facture $S_{AA}(k)$. }
\label{fig:prop-tauqt-mutual-visc-3DKA}
\end{center}
\end{figure}

The breakdown of the SE relation in three dimensions in the Kob-Andersen model has been reported earlier 
\cite{pap:BordatJPCM,thesis:Ashwin,pap:SE-Affouard2009}. 
Our data agree reasonably well with those of previous studies. Here we study the dependence of the SE relation on the spatial 
dimension by considering the following models which are defined in the previous section: (i) 2DKA, (ii) 2DMKA  (iii) 2DR10, 
(iv) 3DKA and (v) 4DKA. According to the SE relation, the quantity $D \eta/T$ should be
independent of the temperature $T$ -- deviations of this ratio from a constant value as the temperature is changed would 
constitute a violation of the SE relation. Since the viscosity $\eta$ is difficult to calculate in a molecular dynamics 
simulation, the temperature dependence of $D \tau_\alpha$ or $D \tau_\alpha/T$ has been considered in most existing 
numerical studies to investigate the validity of the SE relation. These choices correspond to a replacement of 
$\eta$ by $\tau_{\alpha}$ assuming either that $\tau_{\alpha} \propto \eta$ or $\tau_{\alpha} \propto \eta/T$. 
The justification for these substitutions obtain from either the Maxwell relation ($\eta = G_{\infty} \tau$) or the 
relationship between relaxation times and the diffusion coefficient in the diffusive regime 
($\tau^{-1} = D q^2 = \frac{mk_{B} q^2 }{c \pi R}\frac{T}{\eta}$; this argument assumes the validity of the SE relation). 
From various experimental and simulation results, 
\cite{mezei,yamamoto,chong}, $\tau_{\alpha} \propto \eta/T$. Consistently with previous observations, we find that 
$\tau_\alpha \propto \eta/T$ provides a very good description of the low-temperature data over a fairly large range 
of $\eta$ and $\tau_\alpha$, as shown in Fig.\ref{fig:prop-tauqt-mutual-visc-3DKA}, using viscosities calculated from 
the Green-Kubo formula and the Einstein relation for the 3DKA model. We therefore use the following two representations 
to analyze the degree of SE violation from the data.

\begin{enumerate}[label=\roman*, leftmargin=1em]
\item
In a $D_{A}$ {\it vs.} $\tau_{\alpha}$ or $\frac{\eta}{T}$ plot ($D_A$ is the calculated diffusion coefficient of particles 
of type $A$), by fitting the data to a power law of the form $D_{A} \propto \tau_{\alpha}^{-\xi}$, or 
$D_{A}  \propto \left(\frac{\eta}{T}\right)^{-\xi}$ we test if the effective exponent $\xi$ is different from 
unity, the prediction of the SE relation. 
\item
From the $T$ dependence of $D_{A}\tau_{\alpha}$ or $\frac{D_{A}\eta}{T}$, we test if these quantities are constant 
(SE relation obeyed) or become $T$ dependent (breakdown of the SE relation).
\end{enumerate}

\begin{figure}[!h]
\begin{center}
\vspace{7mm}
\includegraphics[width=6.7cm,height=5.2cm]{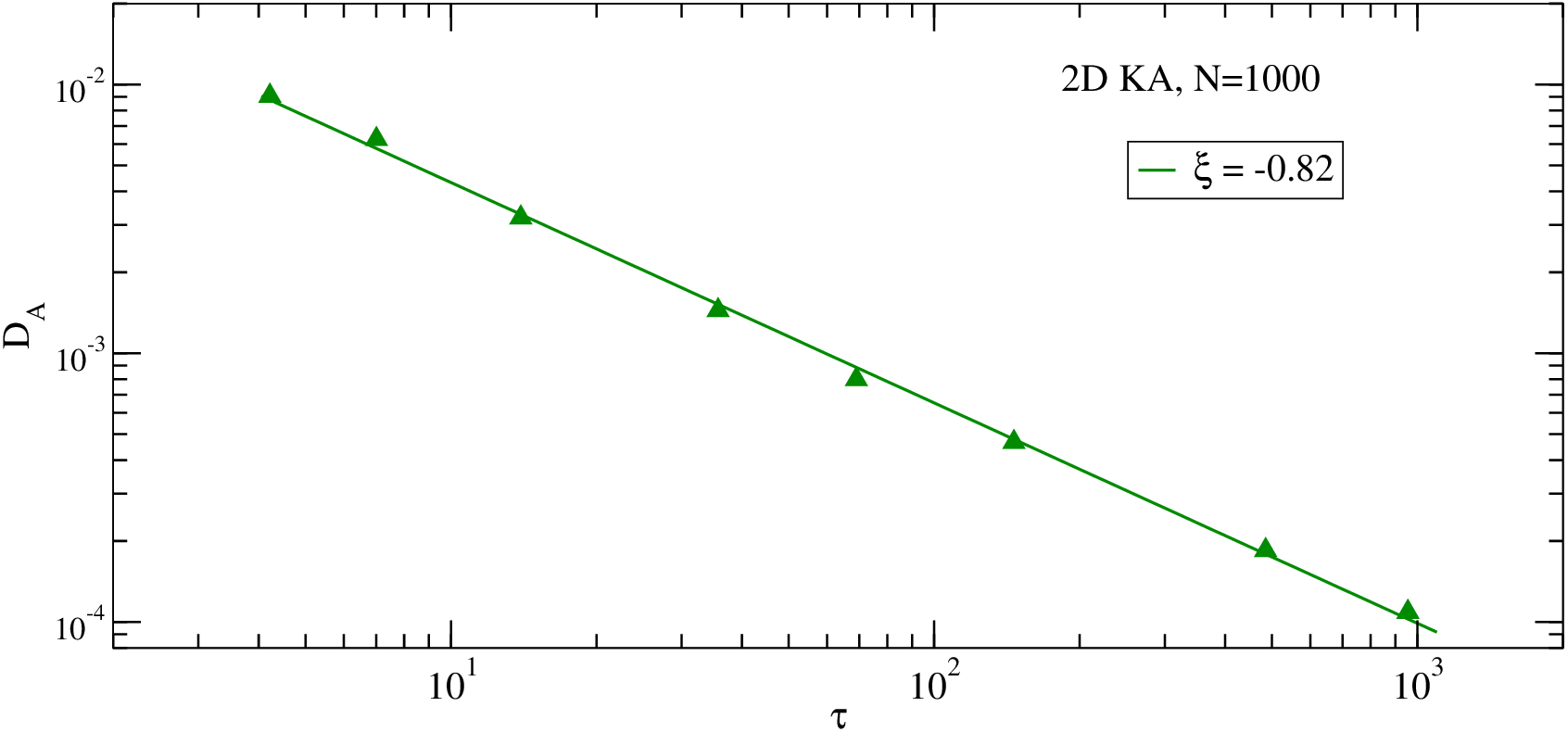} 
\vspace{5mm}
\includegraphics[width=6.8cm,height=5.2cm]{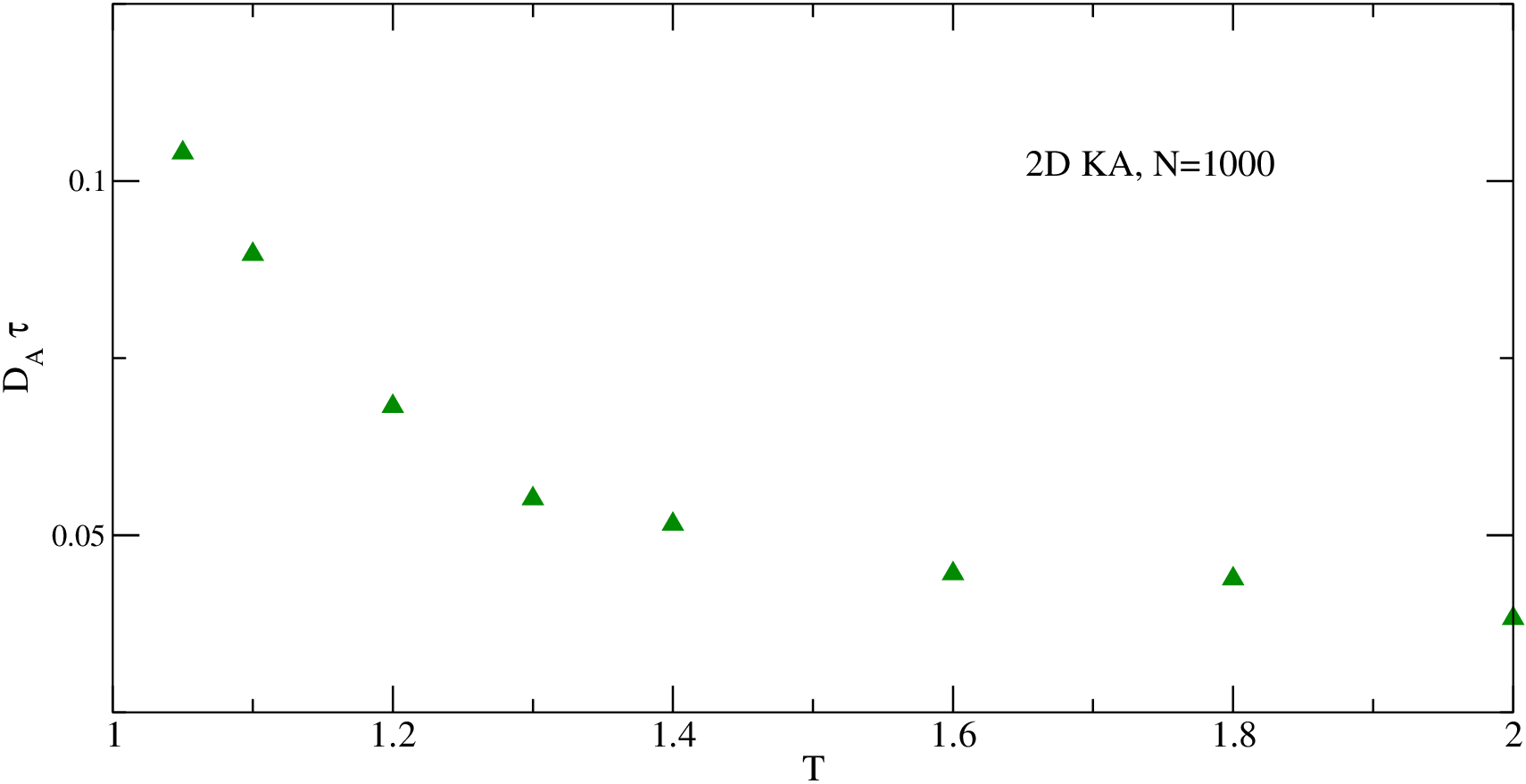} 
\caption{Plot showing the breakdown of the SE relation in the 2D KA model at low temperatures. 
\emph{Top}: $D_{A}$ {\it vs.} $\tau$ plot (Here $\tau$
  is the $\alpha$ relaxation time from the overlap function $q(t)$).
  \emph{Bottom}: $T$ dependence of $D_{A}\tau$.  The data
shown follow a fractional SE relation.}\label{fig:2DKA-N1000-SEdata}
\end{center}
\end{figure}

\begin{figure*}[t!]
\begin{center}
\includegraphics[width=6.7cm,height=5.5cm]{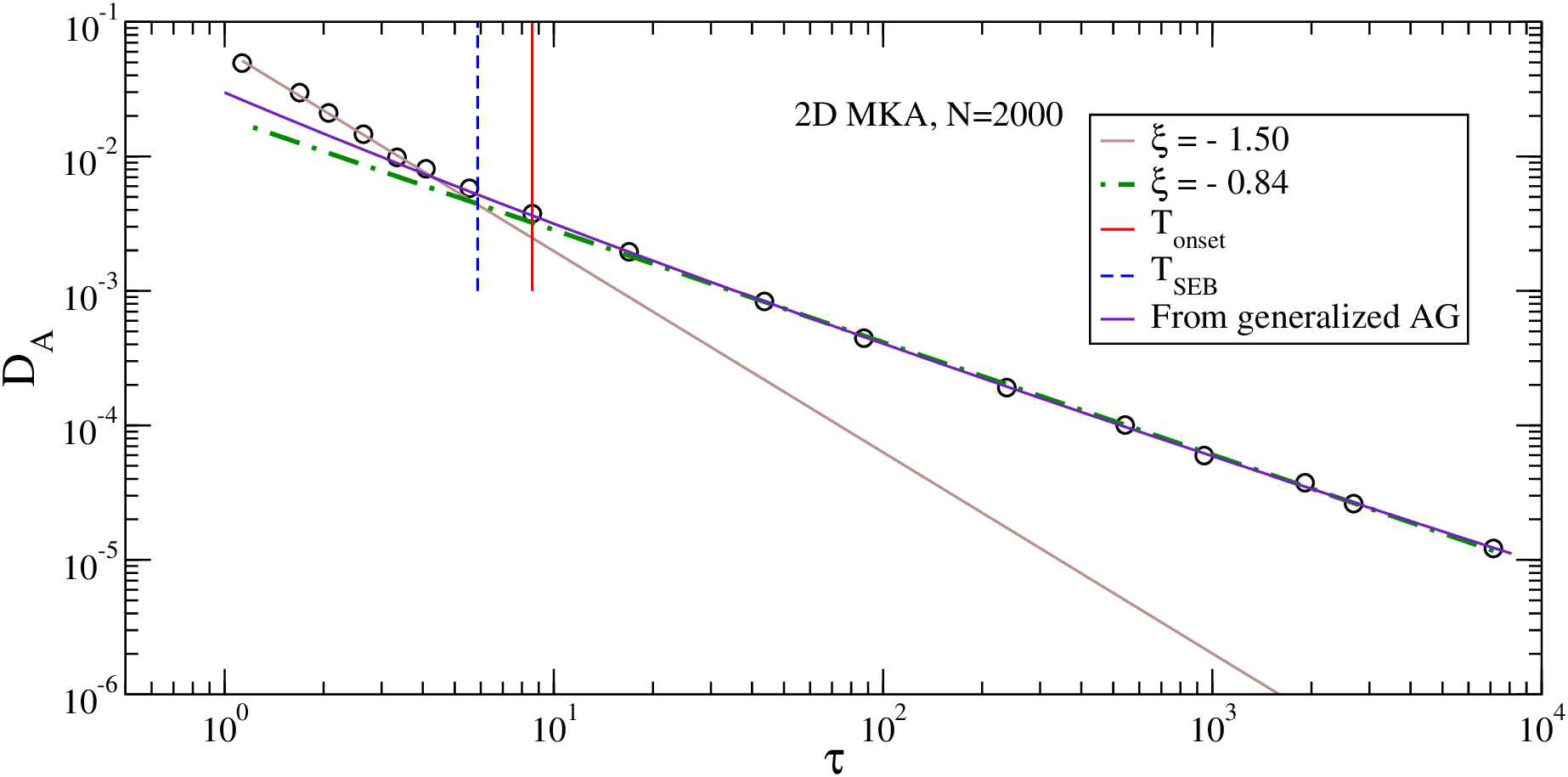} 
\hskip 0.3cm
\includegraphics[width=6.7cm,height=5.5cm]{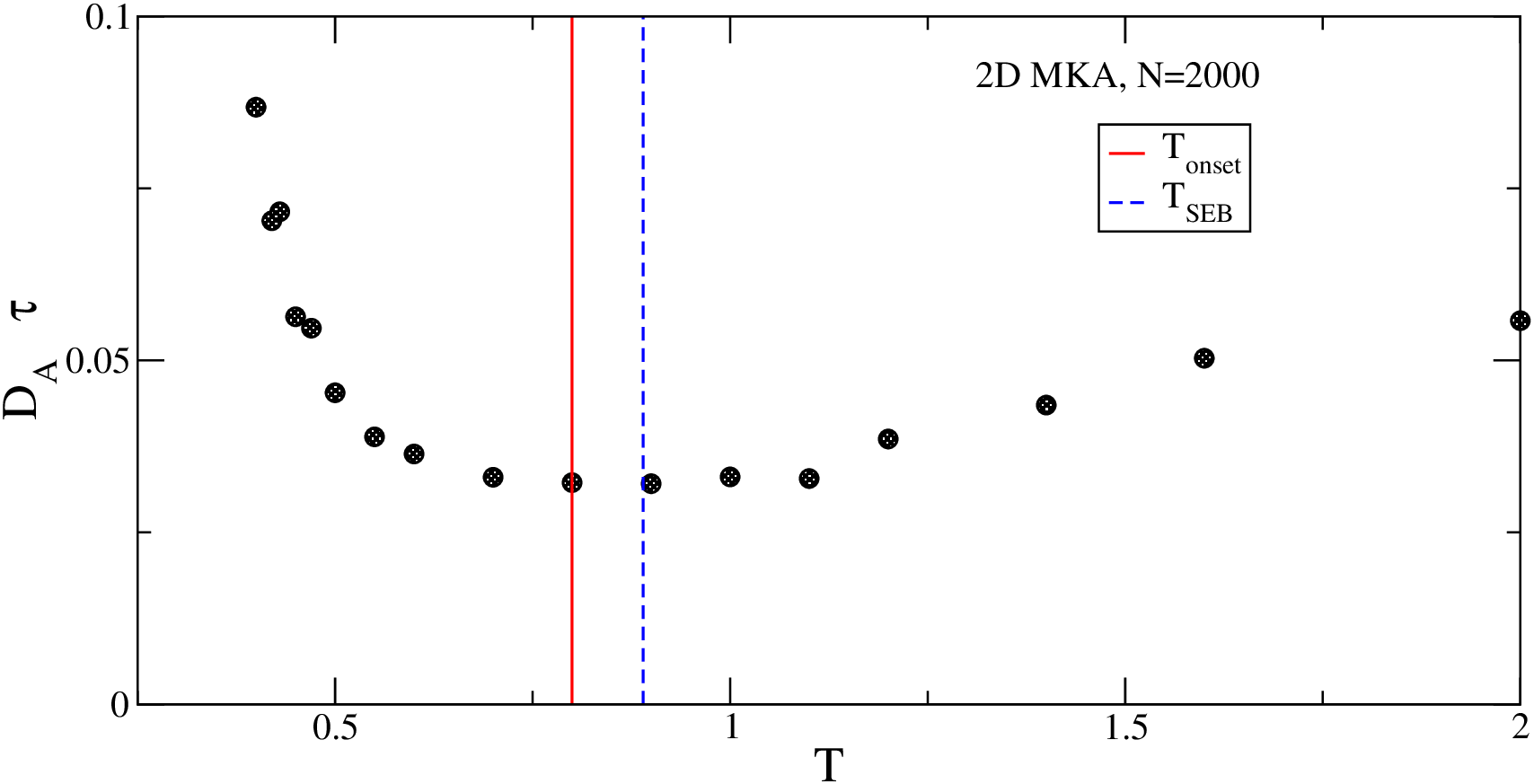} 
\caption{Plots showing the breakdown of the SE relation in the 2DMKA model. 
\emph{Left}: $D_{A}$ {\it vs.} $\tau$ plot (Here $\tau$ is the $\alpha$ relaxation time from the overlap function $q(t)$). 
\emph{Right:} $T$ dependence of $D_{A}\tau$. 
The low T data follow a fractional SE relation. A clear change of exponent occurs at high T in $D_{A}$ {\it vs.} $\tau$ plot, 
although \emph{the high T  exponent is bigger than 1}. The change of slope occurs at a temperature $T_{SEB}$ which is close
to $T_{onset}$. ($T_{SEB}$ estimated as the point of intersection of high T and low T fits; $T_{onset}$ is the onset temperature of slow dynamics.) Also shown is the dependence of $D$ on $\tau$ according to the generalized Adam-Gibbs relation discussed in the text.}\label{fig:2DMKA-N2000-SEdata}
\end{center}

\begin{center}
\includegraphics[width=6.7cm,height=5.9cm]{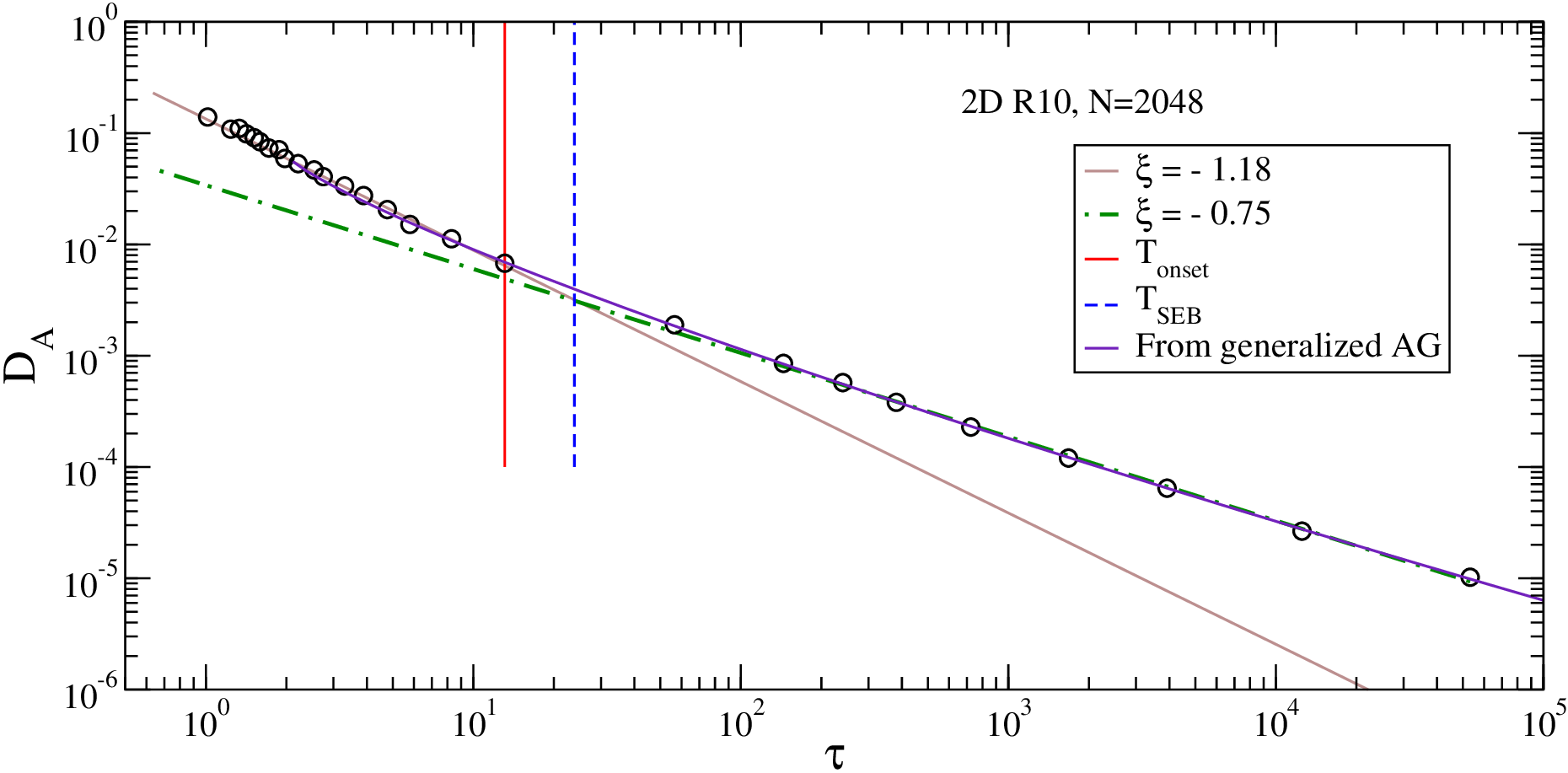} 
\hskip 0.3cm
\includegraphics[width=6.7cm,height=5.7cm]{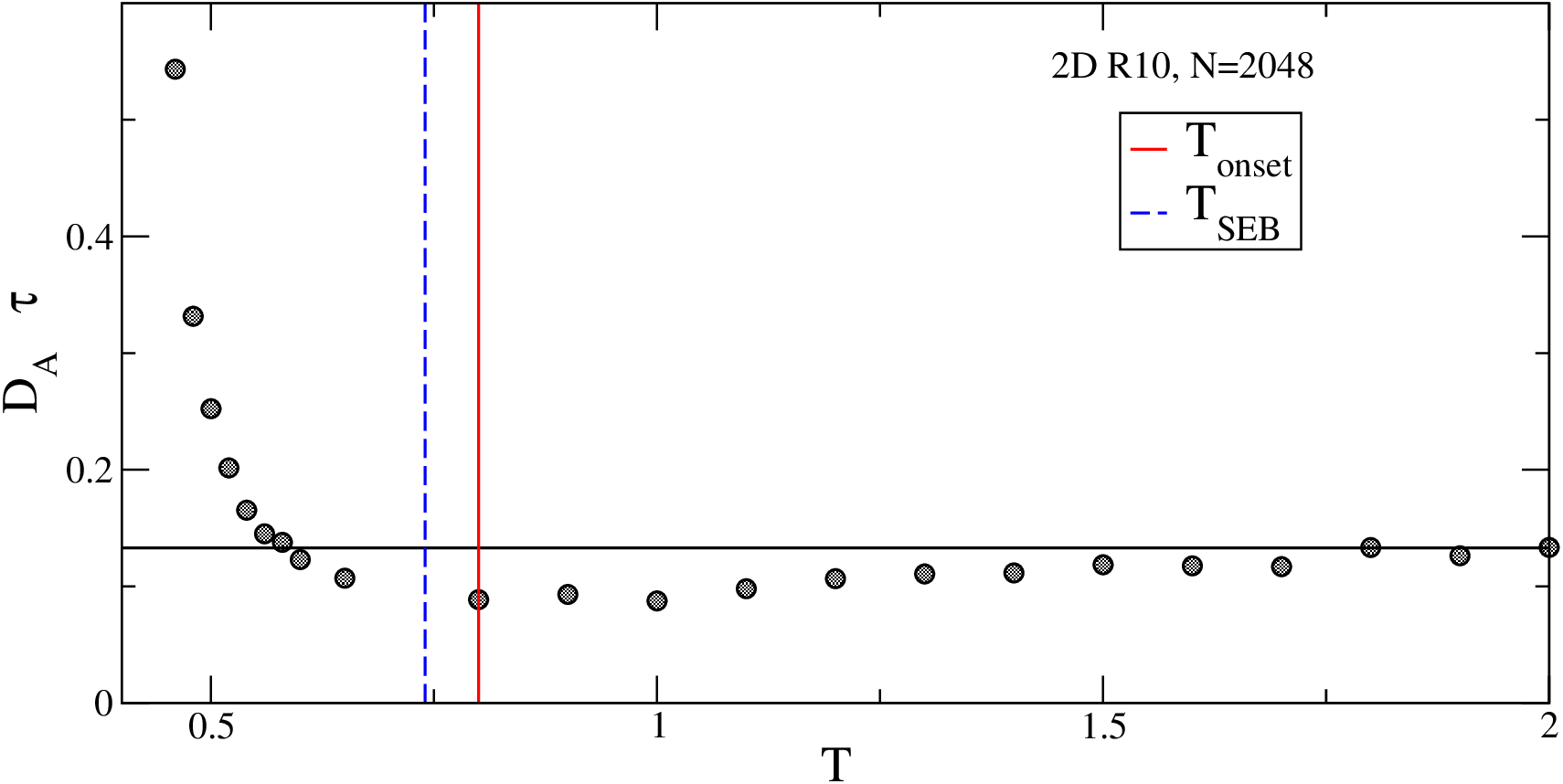} 
\caption{Plots showing the breakdown of the SE relation in the 2DR10
  model.  \emph{Left}: $D_{A}$ {\it vs.} $\tau$ plot (Here $\tau$
  is the $\alpha$ relaxation time from the overlap function $q(t)$).
  \emph{Right:} $T$ dependence of $D_{A}\tau$.  The low T data
  follow a fractional SE relation. A clear change of exponent occurs
  at high T in $D_{A}$ {\it vs.} $\tau$ plot, although \emph{the
    high T exponent is bigger than 1}. The change of slope occurs at a
  temperature $T_{SEB}$ which is close to $T_{onset}$. ($T_{SEB}$ estimated 
as the point of intersection of high T and low T fits; $T_{onset}$ is the onset 
temperature of slow dynamics.) Also shown is the dependence of $D$ on $\tau$ according to the 
generalized Adam-Gibbs relation discussed in the text.}\label{fig:2DR10-N2048-SEdata}
\end{center}
\end{figure*}

\begin{figure*}[!h]
\begin{center}
\includegraphics[width=6.9cm,height=6cm]{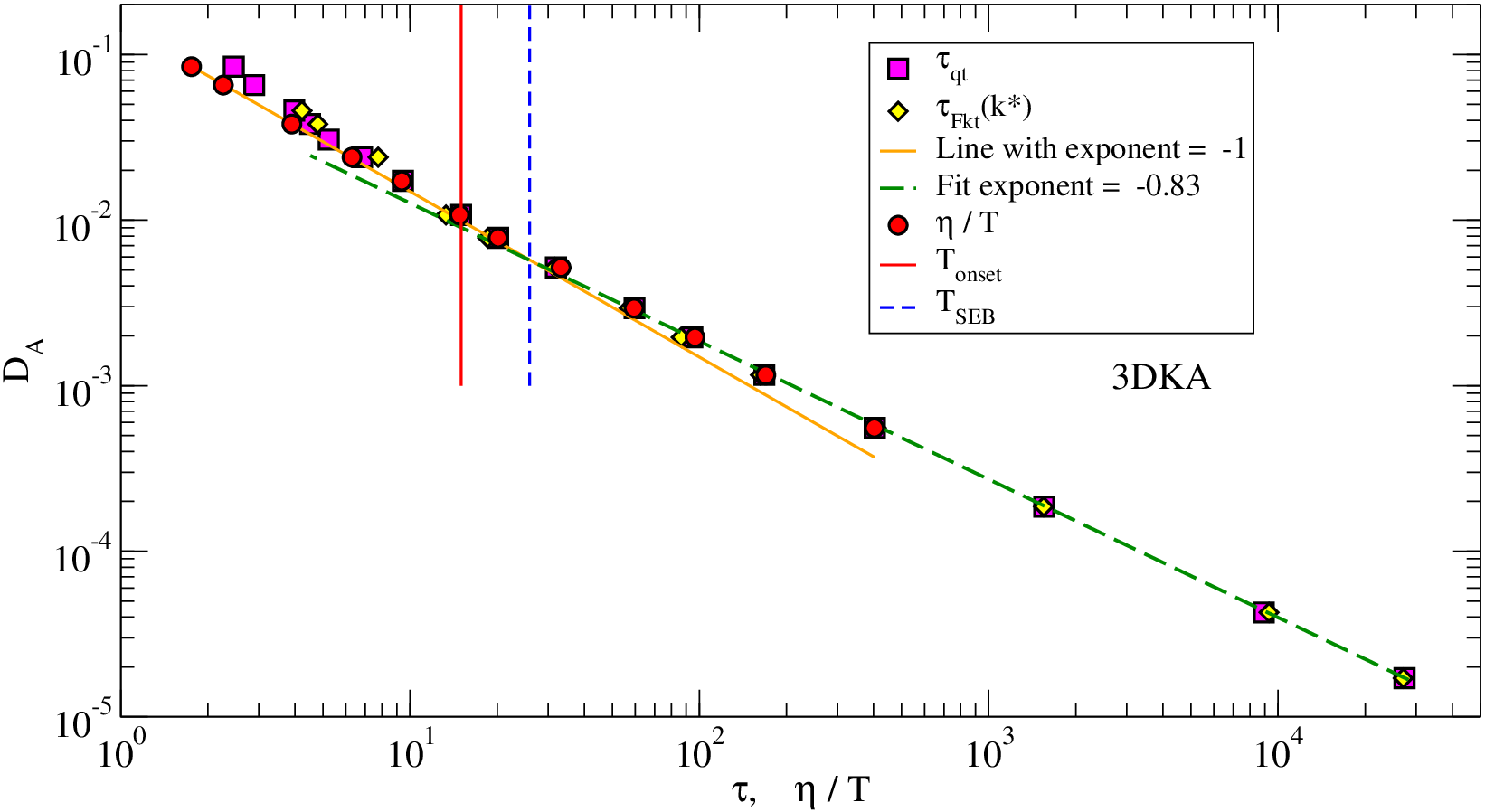} 
\hskip 0.4cm
\includegraphics[width=7.4cm,height=6cm]{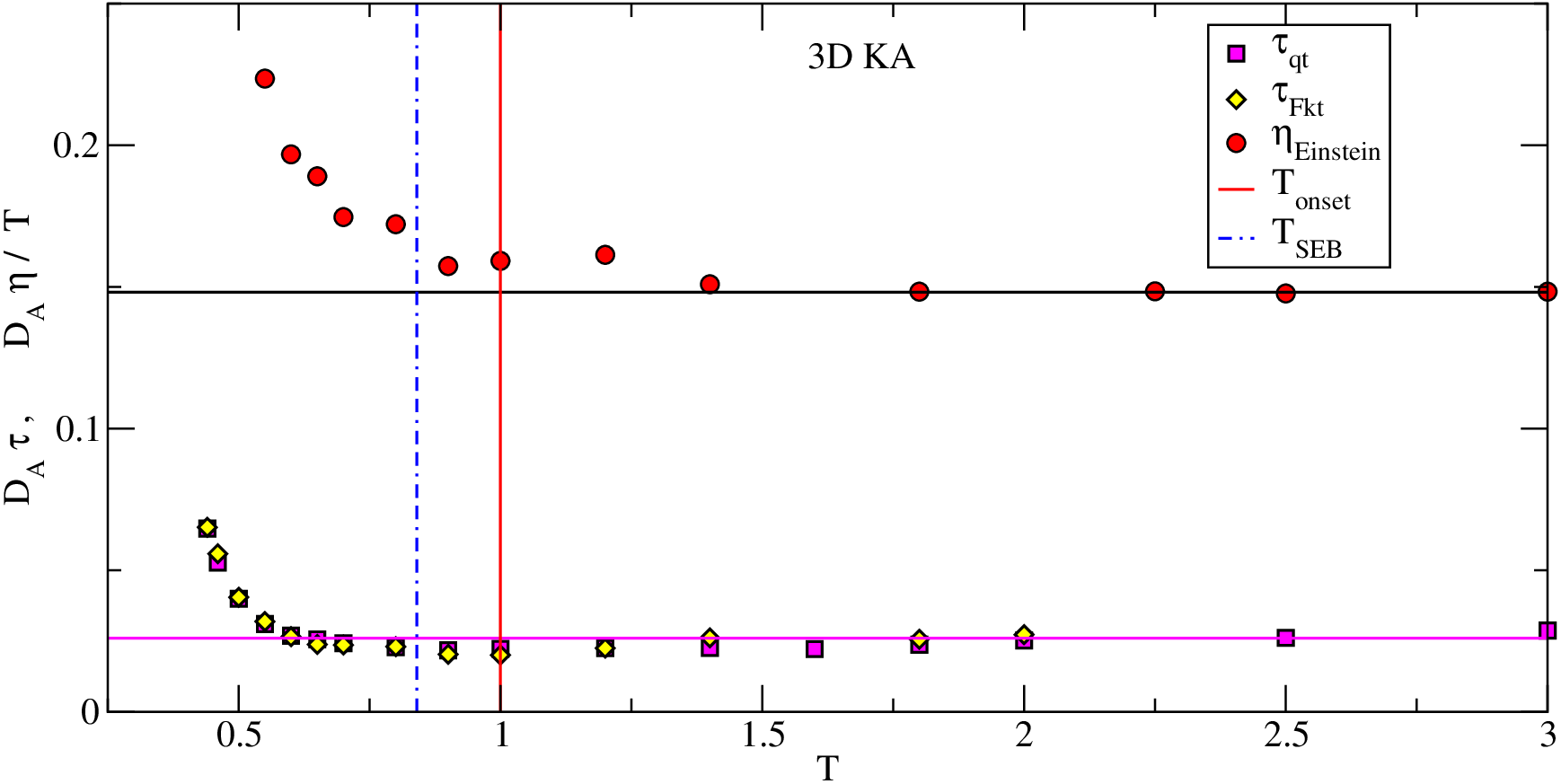} 
\caption{Plots showing the breakdown of the SE relation in the 3DKA
  model. The $\alpha$ relaxation times $\tau$ are computed
  from the (i) overlap function ($\tau_{qt}$) (ii) $F(k,t)$ at the
  peak of $S(k)$ ($\tau_{Fkt}$) These two measures are mutually proportional
  and are used interchangeably.  \emph{Left}: $D_{A}$ {\it vs.}
  $\tau$ and $\frac{\eta}{T}$ plot. $\tau_qt$, $\tau_{Fkt}(k^*)$ are multiplied 
  by constant factors to match all data sets at low temperature. \emph{Right:} $T$
  dependence of $D_{A}\tau$ and $\frac{D_{A}\eta}{T}$.  The
  low T data follow a fractional SE relation. A clear change of
  exponent occurs at high T in the $D_{A}$ {\it vs.} $\tau$
  plot. The high T exponent (=-1) expected from the SE relation is
  obtained from the $D_{A}$ {\it vs.}  $\frac{\eta}{T}$ plot. The SE
  breakdown occurs at a temperature $T_{SEB}$ closer to the onset
  temperature $T_{onset}$ than $T_{c}$ \cite{pap:KOB-MCT}. ($T_{SEB}$ estimated as the
  point of intersection of high T and low T fits; $T_{onset}$ is the
  Arrhenius to non-Arrhenius cross-over temperature; $T_{c}$ is the
  mode coupling transition temperature. All data points shown here are
  at $T > T_{c}$.)}\label{fig:3DKA-N1000-SEdata}
\end{center}


\hskip 0.1cm
\begin{center}
\includegraphics[width=6.9cm,height=6cm]{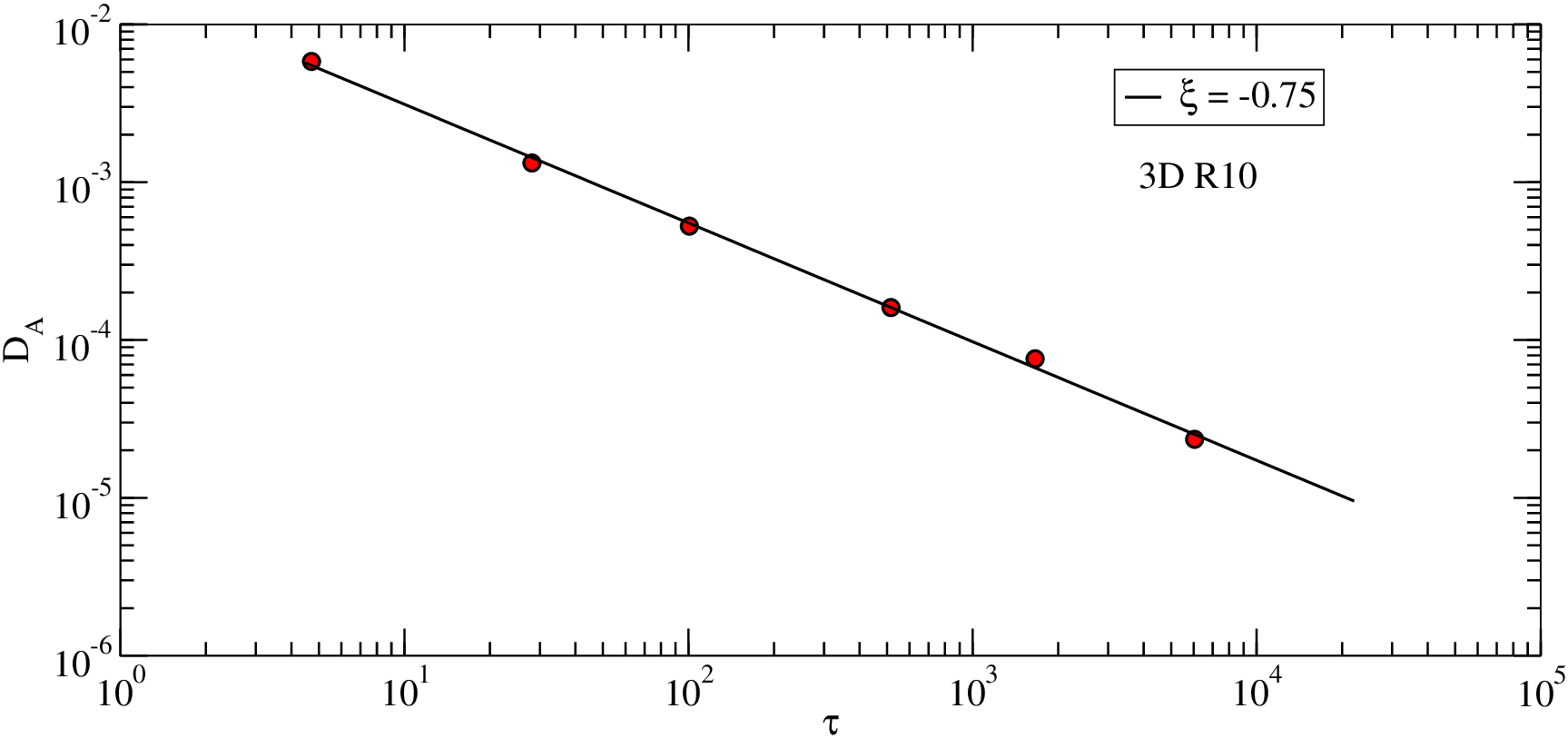} 
\hskip 0.4cm
\includegraphics[width=6.7cm,height=6cm]{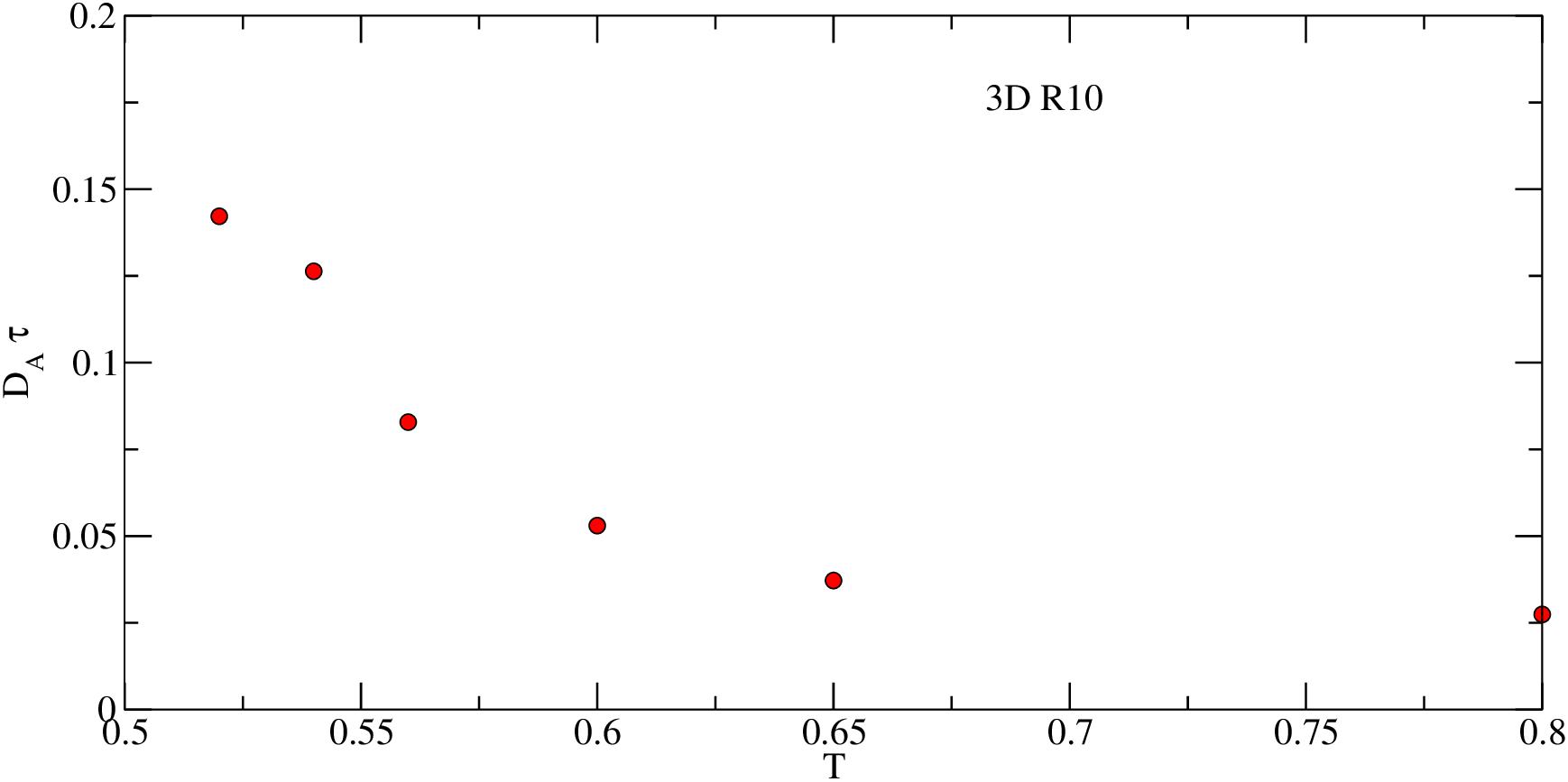} 
\caption{Plot showing the breakdown of the SE relation in the 3D R10 model at low temperatures. \emph{Left}: $D_{A}$ {\it vs.} $\tau$ plot (Here $\tau$
  is the $\alpha$ relaxation time from the overlap function $q(t)$).
  \emph{Right:} $T$ dependence of $D_{A}\tau$.  The data
shown follow a fractional SE relation.}\label{fig:3DR10-N10000-SEdata}
\end{center}
\end{figure*}

\begin{figure*}[!h]
\begin{center}
\includegraphics[width=6.4cm,height=5.5cm]{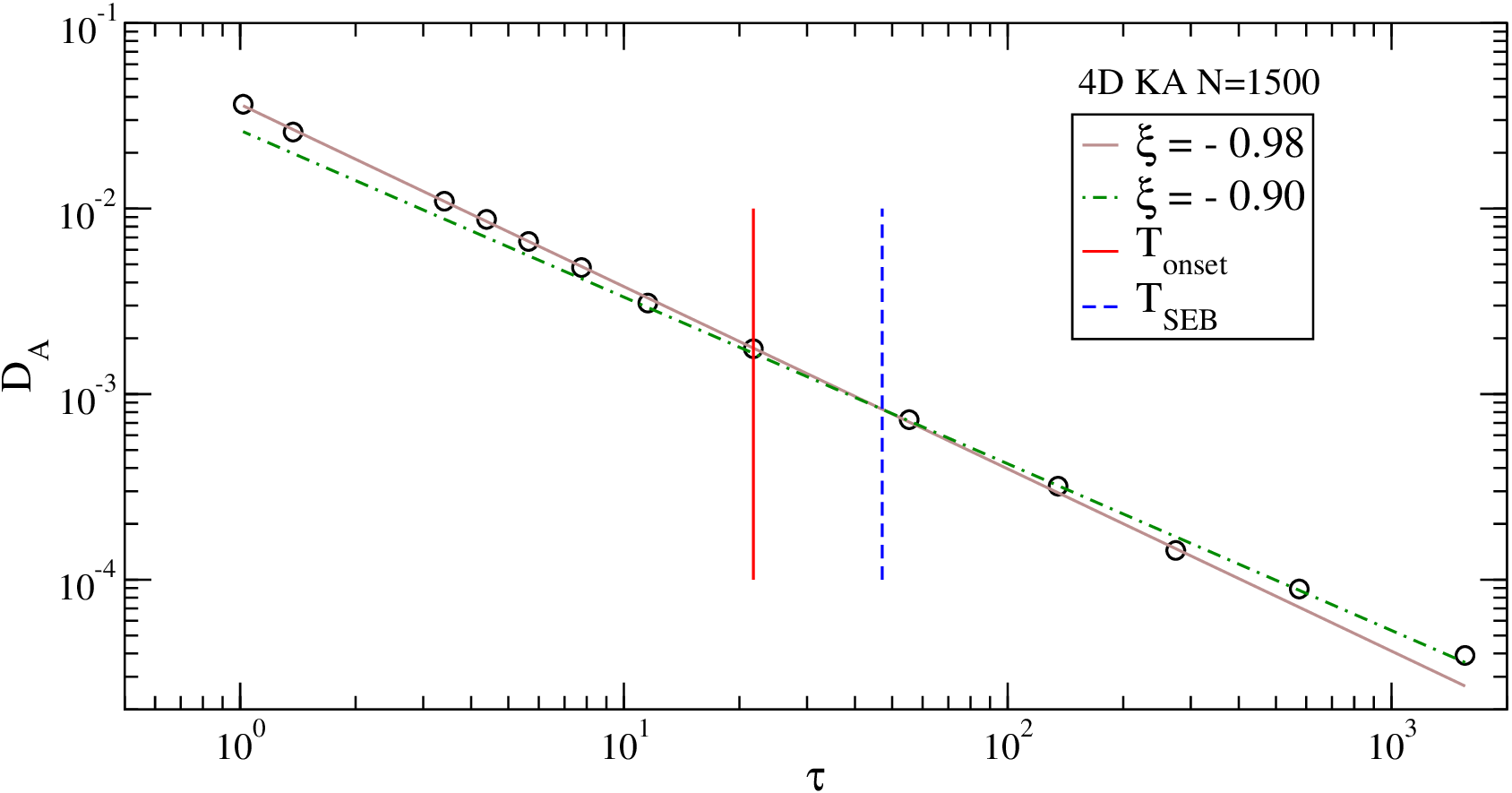} 
\hskip 0.5cm
\includegraphics[width=6.4cm,height=5.5cm]{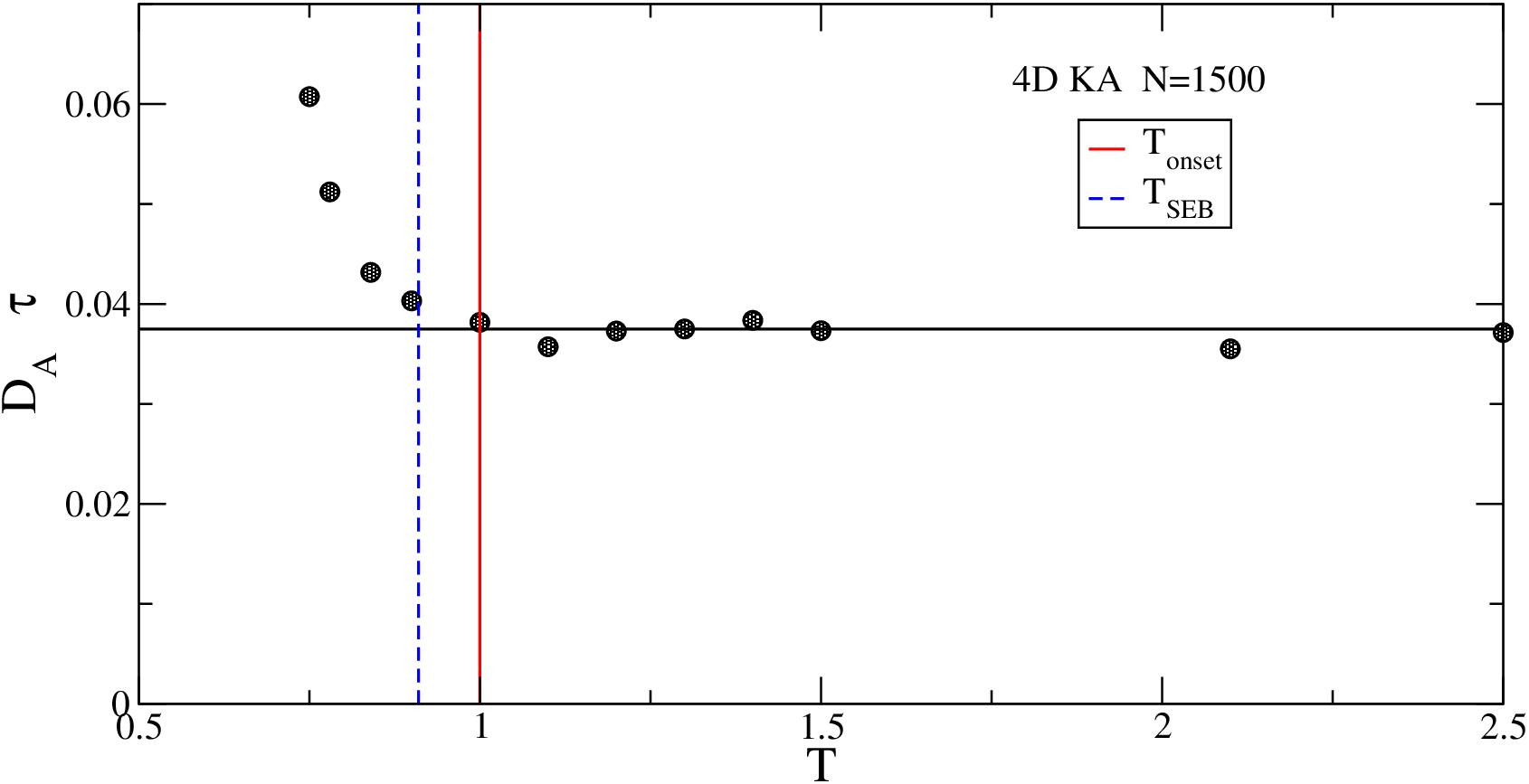} 
\caption{Plots showing the breakdown of the SE relation in the 4DKA
  model.  \emph{Left}: $D_{A}$ {\it vs.} $\tau$ plot ($\tau$ here
  is the $\alpha$ relaxation time from the overlap function $q(t)$).
  \emph{Right:} $T$ dependence of $D_{A}\tau$.  The low T data
  follow a fractional SE relation. A clear change of exponent occurs
  at high T in $D_{A}$ {\it vs.} $\tau$ plot. The high T exponent
  (=-1) expected for a homogeneous (Gaussian distribution of particle
  displacements) system is obtained from $D_{A}$ {\it vs.} $\tau$
  plot. The SE breakdown occurs at a temperature $T_{SEB}$ closer to
  the onset temperature $T_{onset}$ than $T_{c}$. ($T_{SEB}$ estimated
  as the point of intersection of high T and low T fits; $T_{onset}$
  is the Arrhenius to non-Arrhenius cross-over temperature.)}\label{fig:4DKA-N1500-SEdata}
\end{center}
\begin{center}
\includegraphics[width=6.7cm,height=5.5cm]{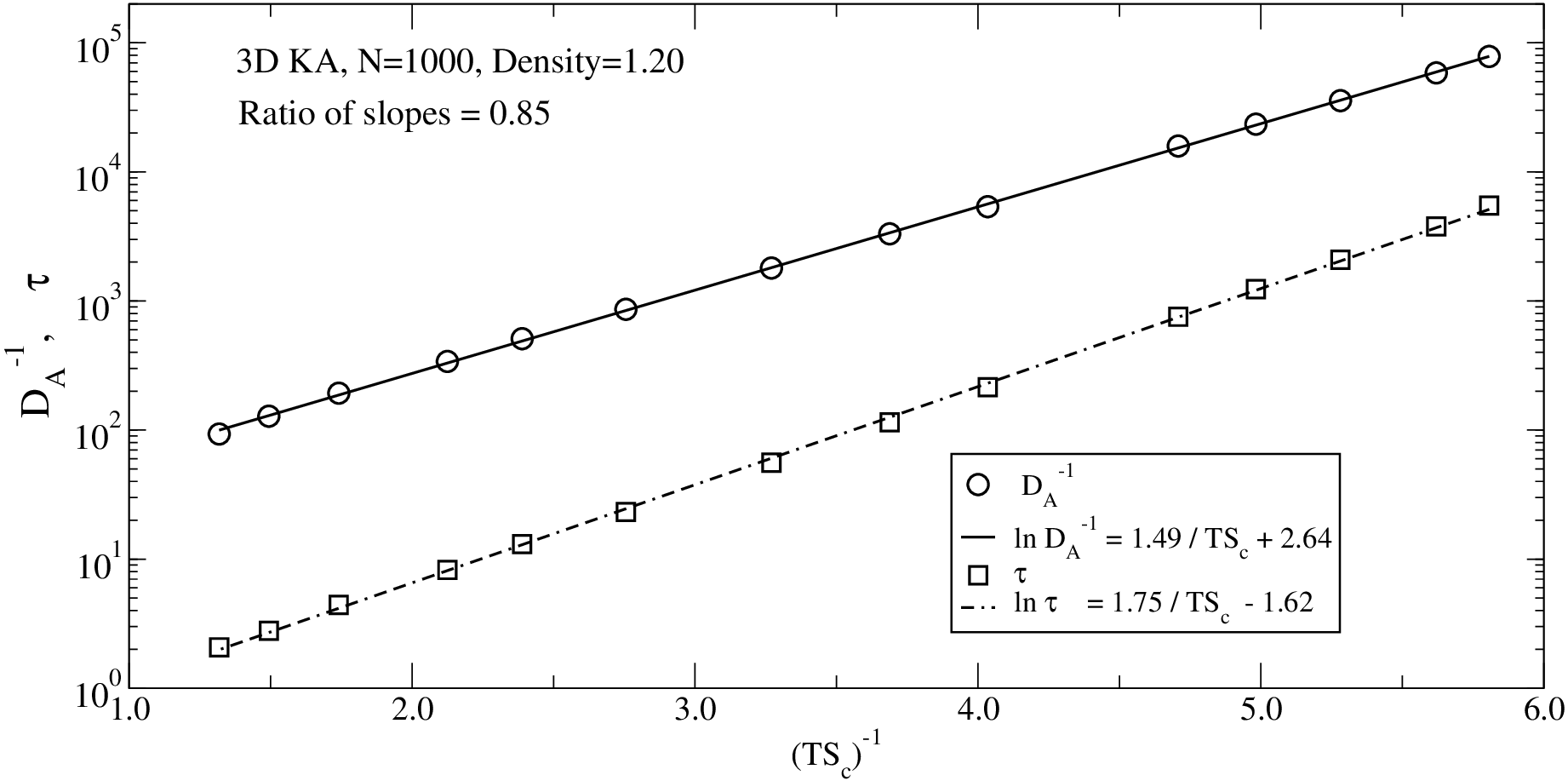} 
\hskip 0.5cm
\includegraphics[width=6.7cm,height=5.5cm]{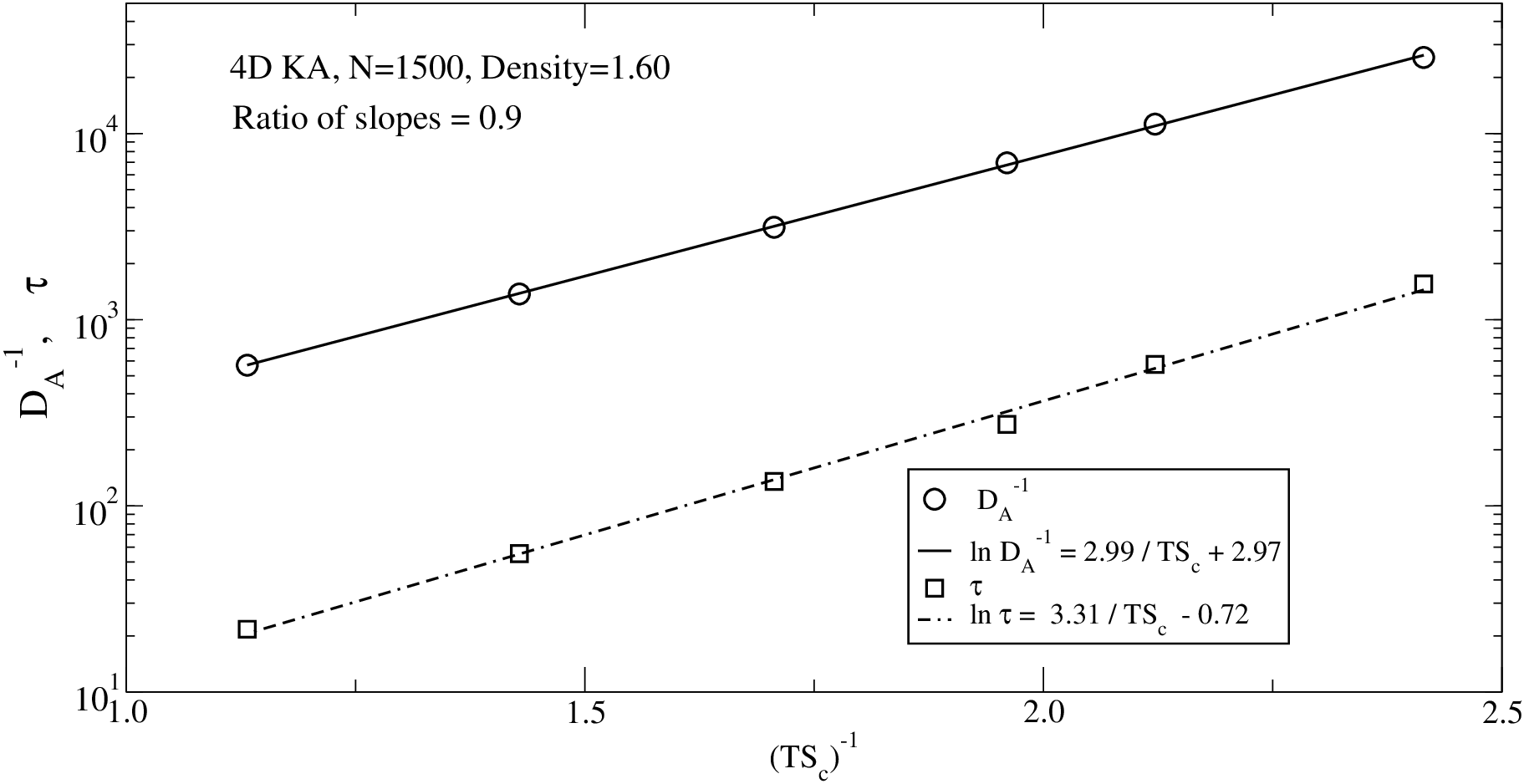} 
\caption{\emph{Left}: The Adam Gibbs (AG) relation in the 3DKA model using as the dynamical quantities $D_{A}$, $\tau$. Here $\tau$ is the $\alpha$ relaxation time from the overlap function $q(t)$. 
\emph{Right}: The AG relation in the 4DKA model using as the dynamical quantities $D_{A}$, $\tau$. 
The slopes are different for $D_{A}$ and $\tau$ indicating that the diffusion coefficient has a 
\emph{different} dependence on the configuration entropy than the $\alpha$ relaxation time. The fractional 
SE exponent at low $T$ can be interpreted as the ratio of the slopes (Table \ref{tab:exponents}). } 
\label{fig:xi-from-AG-3DKA-4DKA}
\end{center}
\begin{center}
\includegraphics[width=6.9cm,height=5.5cm]{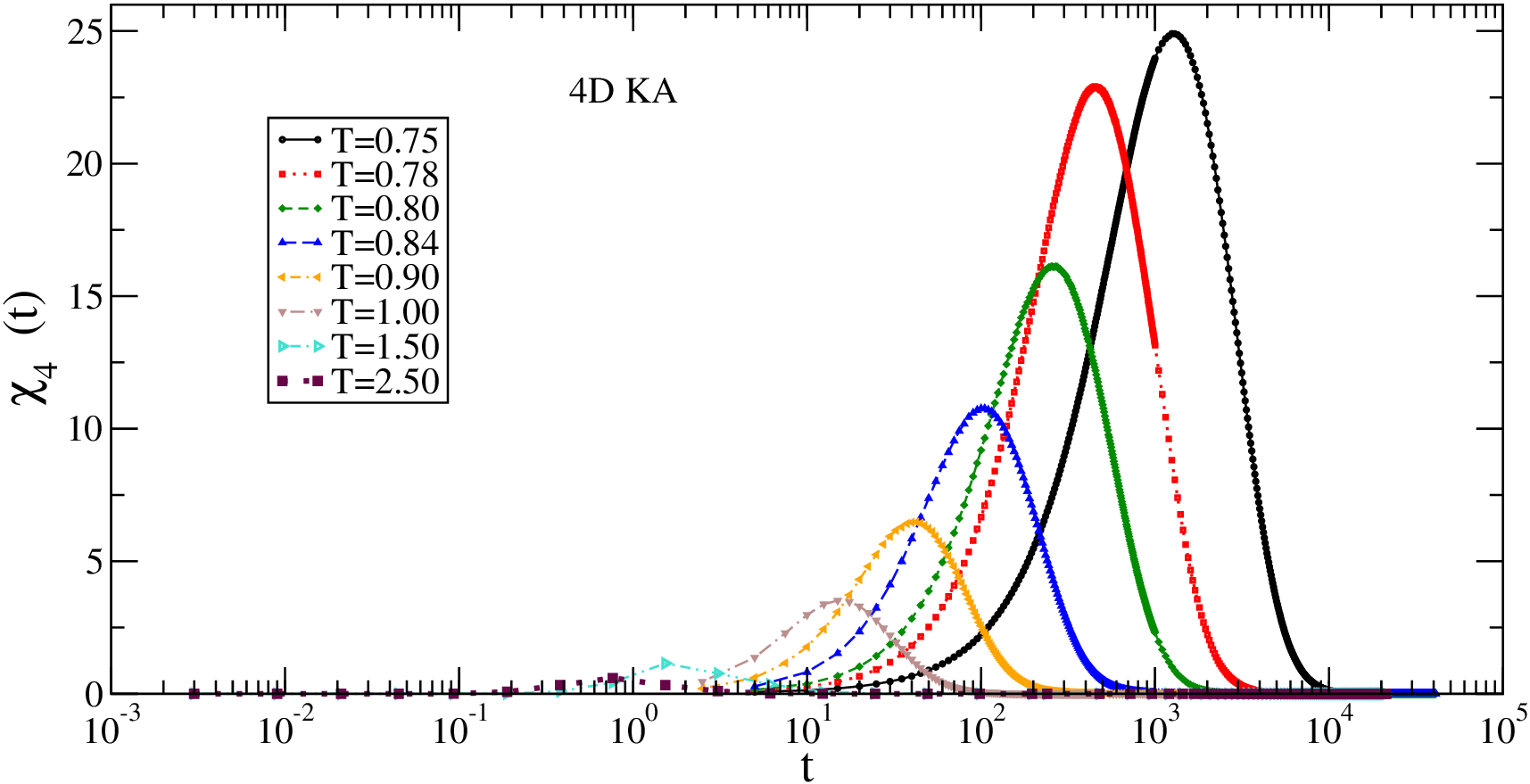} 
\hskip 0.3cm
\includegraphics[width=6.9cm,height=5.5cm]{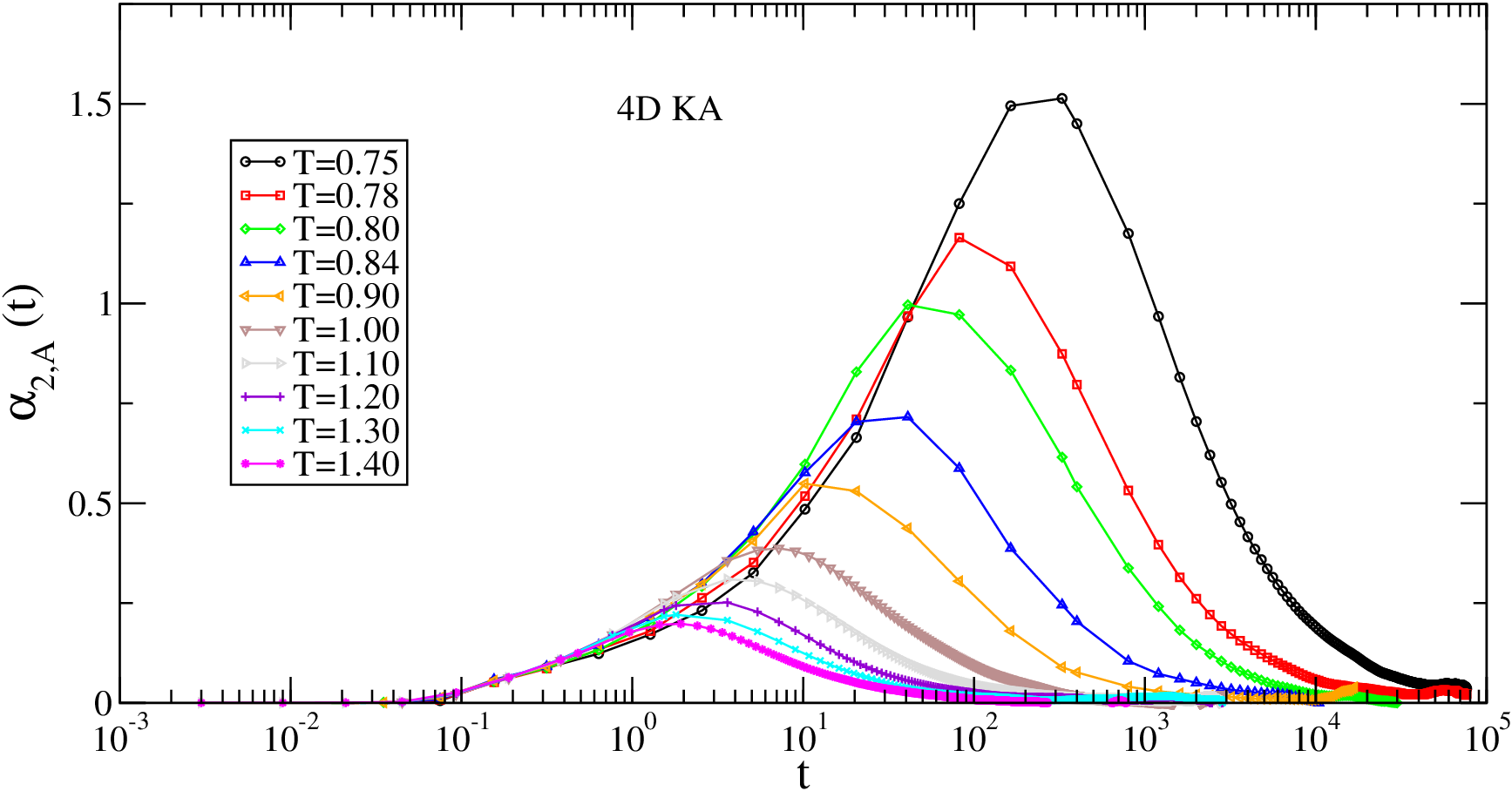} 
\caption{The time dependences of the dynamical susceptibility $\chi_{4}(t)$ and 
of the non-Gaussian parameter $\alpha_{2,A}(t)$ in the 4DKA model. Peak values $\chi_{4}^{peak}$ and $\alpha_{2,A}^{peak}$ extracted from such time dependence are described further below.}\label{fig:alpha2-chi4-vst-4DKA}
\end{center}
\end{figure*}

The breakdown of the SE relation in 2, 3 and 4 dimensions is shown in
Figs. \ref{fig:2DKA-N1000-SEdata} (2DKA), \ref{fig:2DMKA-N2000-SEdata}
(2DMKA), \ref{fig:2DR10-N2048-SEdata} (2DR10),
\ref{fig:3DKA-N1000-SEdata}(3DKA), \ref{fig:3DR10-N10000-SEdata}(3DR10)
and \ref{fig:4DKA-N1500-SEdata} (4DKA). In most of the cases
(excepting 2DKA and 3DR10) we show data ranging from high
temperatures, well above the onset temperature $T_{onset}$, to well
below $T_{onset}$. A change from Arrhenius to super-Arrhenius
temperature dependence of relaxation times defines the onset
temperature of slow dynamics
\cite{pap:AG-Sastry-Deb-Still,pap:Sastry-pcc,pap:Sastry-physA} $T_{onset}$. The onset
temperature is close in all cases to the temperature at which one observes the
breakdown of the SE relation, $T_{SEB}$ as described below. All the simulations reported 
here are performed at temperatures above the mode coupling temperatures $T_{c}$ for 
the respective models, and the  $T_{SEB}$ we obtain are well above the $T_{c}$ values. 

one below the other in single column instead of messing with two column figures. 

From the $D_{A}$ {\it vs.} $\tau_{\alpha}$ or $\frac{\eta}{T}$ plots,
we see that in \emph{all} spatial dimensions, the low $T$ data follow
a fractional power-law relation, indicating a breakdown of the SE
relation.  For the same model (KA) in different dimensions the
power-law exponent $\xi$ is closer to $1$ in higher dimensions,
indicating that the SE breakdown is weaker in higher dimensions
\cite{pap:Eaves-Reichmann,pap:Charbonneau-etal}. The difference in
the exponent in two and three dimensions however is very small and
essentially negligible. Similarly for the R10 model, the exponent is
marginally, but negligibly, higher in three dimensions. The MKA
model in two dimensions has an exponent that is very close to the KA
model in the same dimension.  All the model results taken together,
the breakdown exponent is different for different models in the same
dimension. 

Further, as the temperature increases, there is a clear change of the
exponent value in $D_{A}$ {\it vs.} $\tau_{\alpha}$ or
$\frac{\eta}{T}$, plots indicating a \emph{qualitative} difference
between high $T$ and low $T$ behaviours. Surprisingly, in two
dimensions, in both the models for which high $T$ data are shown
(2DMKA and 2DR10), the high $T$ exponent from $D_{A}$ {\it vs.}
$\tau_{\alpha}$ plots is \emph{higher than 1}. However, in 3
dimensions, the SE relation $D_{A} \propto \frac{\eta}{T}$ is
recovered at high $T$ as expected.  Finally, as we go to still higher
dimension ($D=4$), the $D_{A}$ {\it vs.} $\tau_{\alpha}$ plot shows
the expected relation ($D_{A} \propto \tau_{\alpha}^{-1}$) at high
$T$.

The above correlation between $D_{A}$ and $\tau_{\alpha}$ is also
reflected in the $T$ dependence of $D_{A}\tau_{\alpha}$. Since the
diffusion coefficient decreases but the relaxation time increases as
the temperature decreases, there is a competition between two opposing
effects.  The respective rates of increase and decrease with $T$
exactly cancel each other \emph{only} if the exponent is 1. Since in
2D, the exponent is never 1, the quantity $D_{A}\tau_{\alpha}$ goes
through a minimum and approaches a constant value only at very high
temperatures.  In 3D and 4D, $D_{A}\tau_{\alpha}$ becomes constant at
high $T$ as expected and it \emph{increases} as $T$ decreases at low
$T$, indicating a breakdown of the SE relation.

We have also considered the behaviour of $D_{A} / T$ {\it vs.} $\tau_{\alpha}$ and the $T$ dependence of 
$\frac{D_{A}\tau_{\alpha}}{T}$. This corresponds to assuming that $\tau_\alpha \propto \eta$.  
The $D_{A} / T$ {\it vs.} $\tau_{\alpha}$ plots show a fractional SE relation at low $T$ in 
\emph{all   dimensions}.  The values of the breakdown exponents from the $D_{A} / T$ {\it vs.} 
$\tau_{\alpha}$ plot are slightly different from those obtained from the $D_{A}$ {\it vs.} $\tau_{\alpha}$ plots. 
However, they show the same trend: the magnitude of the exponent is closer to 1 at higher dimensions, 
indicating that the SE breakdown is weaker at higher dimensions.  The estimates of the breakdown exponents 
are summarized in Table \ref{tab:exponents}.
\begin{table}[!h]
\begin{center}
\caption{Estimates of the magnitude of the SE breakdown exponents in different spatial dimensions $D$. \emph{Notations:} 
(a) $\xi^{SE}$ = SE breakdown exponent obtained from $D_{A}$ {\it vs.} $\tau_{\alpha}$ or $\frac{\eta}{T}$ plots; 
(b) $\xi^{AG}$ = ratio of slopes from AG plots using $D_{A}$ and $\tau_{\alpha}$; 
High T exponents are obtained from $D_{A}$ {\it vs.} $\tau_{\alpha}$ plots. }
\vspace{5mm}
\begin{tabular}{|c|c|c|c|c|}
\hline
\multirow{3}{*}{$D$} & \multirow{3}{*}{Model} & \multicolumn{2}{|c|}{Low T Exponents} & \multirow{3}{*}{High $T$ exponents } \\ 
\cline{3-4}
    &       &               &                                    &                   \\
    &       & $\xi^{SE}$  &  \multicolumn{1}{c|}{$\xi^{AG}$}       &                   \\
\hline
2   & 2DR10 &    0.75       &  \multicolumn{1}{c|}{-}            &      1.18         \\
\hline                                                           
2   & 2DKA  &    0.82       &  -                                 &        -          \\
\hline
2   & 2DMKA &    0.84       &  \multicolumn{1}{c|}{-}            &      1.50         \\
\hline
3   & 3DR10 &    0.752      &          -                         &        -          \\
\hline
3   & 3DKA  &    0.83       &  \multicolumn{1}{c|}{0.85}         &    $\approx 1$    \\
\hline                                                           
4   & 4DKA  &    0.90       &  \multicolumn{1}{c|}{0.90}         &  $0.98 \approx 1$ \\
\hline
\end{tabular}
\label{tab:exponents}
\end{center}
\end{table}


Figs. \ref{fig:2DMKA-N2000-SEdata}, \ref{fig:2DR10-N2048-SEdata},
\ref{fig:3DKA-N1000-SEdata} and \ref{fig:4DKA-N1500-SEdata} also show
that the temperature of SE breakdown ($T_{SEB}$), estimated as the
point of interaction of the low $T$ and the high $T$ fits in $D_{A}$
{\it vs.} $\tau_{\alpha}$ or $\frac{\eta}{T}$ plots, is close to the
Arrhenius to non-Arrhenius cross-over temperature $T_{onset}$ in all
dimensions. Thus, the SE breakdown occurs
at a temperature that is much higher than the divergence temperature
$T_{c}$ of the mode-coupling theory estimated from a power-law fit of
the $T$-dependence of the relaxation time.

The fractional SE behaviour at low $T$ can be rationalized by
considering the \emph{different} dependence of the diffusion
coefficient and the $\alpha$ relaxation time on the configuration
entropy ($S_{c}$). The Adam-Gibbs (AG) relation
$X=X_{0}\exp(\frac{A_{x}}{TS_{c}})$, if it is valid, provides a way to
test this hypothesis quantitatively ($X$ is $\tau_{\alpha}$ or
$(D_{A})^{-1}$
in the present study). Fig. \ref{fig:xi-from-AG-3DKA-4DKA} shows that the AG relation is valid in 
the 3DKA model (top row) and the 4DKA model (bottom row). We see that the slope of the $(D_{A})^{-1}$ 
{\it vs.} $(TS_{c})^{-1}$ plot 
is \emph{different} from that of the $\tau_{\alpha}$ {\it vs.} $(TS_{c})^{-1}$ plot. Table \ref{tab:exponents} shows that the 
observed fractional SE exponent at low $T$ can be interpreted as the ratio of the slopes in the AG plots in 
Fig. \ref{fig:xi-from-AG-3DKA-4DKA}.

In the two dimensional models one sees a deviation from the AG relation at low temperatures. But the behaviour can still 
be rationalized by considering the generalized Adam-Gibbs relation that is a good description of the data 
\cite{pap:AG-Sengupta}. Thus, we can write 

\begin{eqnarray}
\ln D_{A} &=& \ln D_{0} - \left(\frac{A_{D}}{TS_{c}}\right)^{\alpha_{D}} \nonumber\\
\ln \tau &=& \ln \tau_{0} + \left(\frac{A_{\tau}}{TS_{c}}\right)^{\alpha_{\tau}} \nonumber\\
\ln  D_{A} &=& \ln D_{0} - \left(\frac{A_{D}}{A_{\tau}}\right)^{\alpha_{D}} \left[ \ln \frac{\tau}{\tau_{0}} \right] ^ {r} \nonumber\\
r&\equiv&  \frac{\alpha_{D}}{\alpha_{\tau}} \label{eq:genag} 
\end{eqnarray}

Eq. \ref{eq:genag} provides the relationship between $D$ and $\tau$
based on each obeying a generalized Adam-Gibbs relation, which are
shows for 2DMKA and 2DR10 models in
Figs. \ref{fig:2DMKA-N2000-SEdata}, \ref{fig:2DR10-N2048-SEdata}. It
is seen that indeed the low temperature data as well as some part of
the high temperature data are well described by this form. It must be
noted that the Eq. \ref{eq:genag} is {\it not} a fractional SE
relation, and thus, the description of the data for these two
dimensional systems through a breakdown exponent is an approximation.

To summarize, results presented here show that: (i) the SE breakdown
in weaker in four dimensions than in three dimensions which is
consistent with earlier works
\cite{pap:Eaves-Reichmann,pap:Charbonneau-etal}. (ii) The breakdown
exponent can be rationalized from the different scaling of the
diffusion coefficient and the relaxation time with the configurational
entropy, either via the AG relation (three and four dimensions) or a
generalized AG relation (two dimensions). (iii) The behaviour in two
dimensions is more complicated, displaying no temperature regime where
the SE relation is valid, consistently with earlier work
\cite{pap:Perera-Harrowell}. (iv) The SE breakdown temperature
$T_{SEB}$ is, in all cases, very close to the onset temperature
$T_{onset}$. 

\begin{figure*}[t!]
\begin{center}
\includegraphics[width=6.7cm,height=5.5cm]{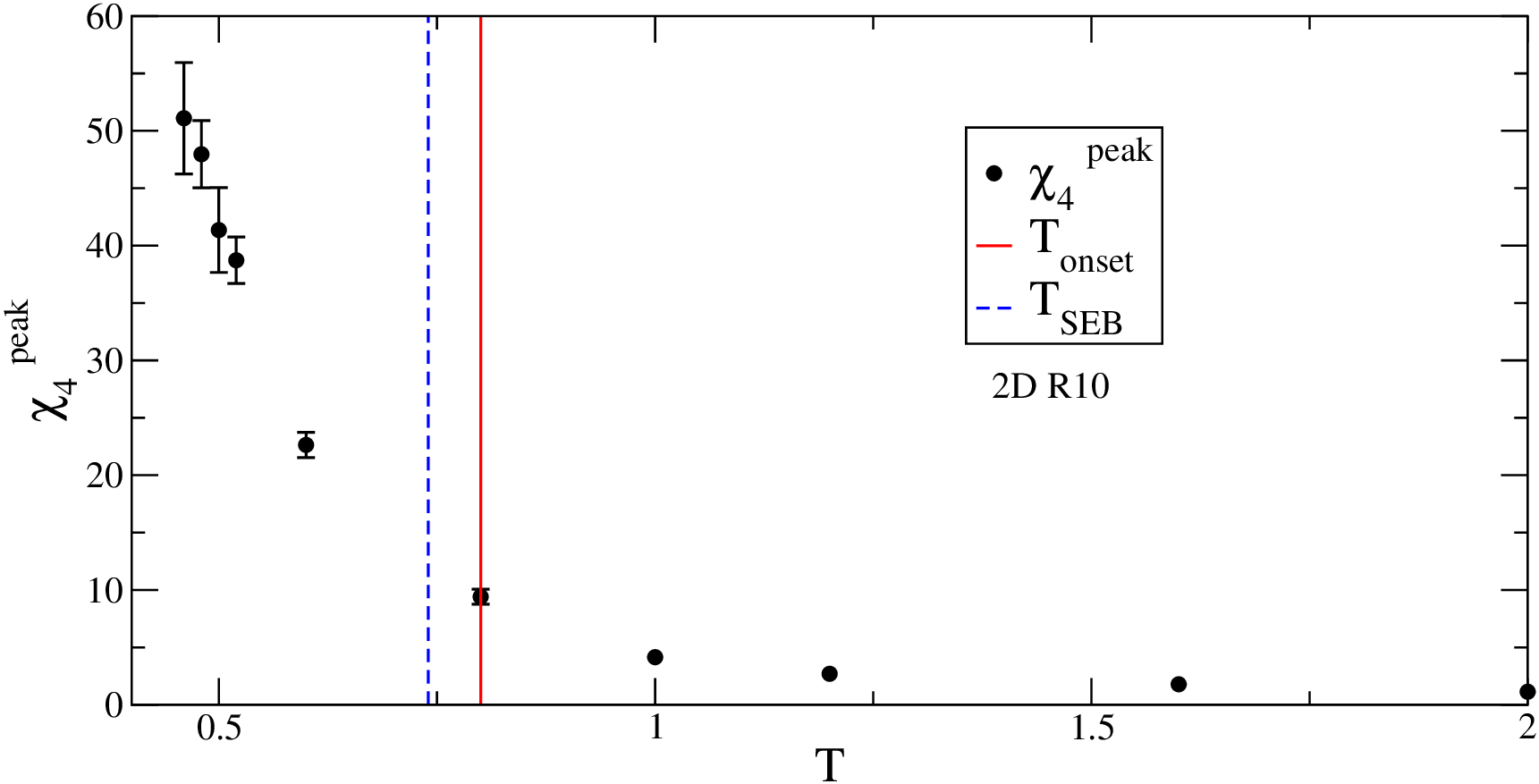} 
\hskip 0.5cm
\includegraphics[width=6.7cm,height=5.5cm]{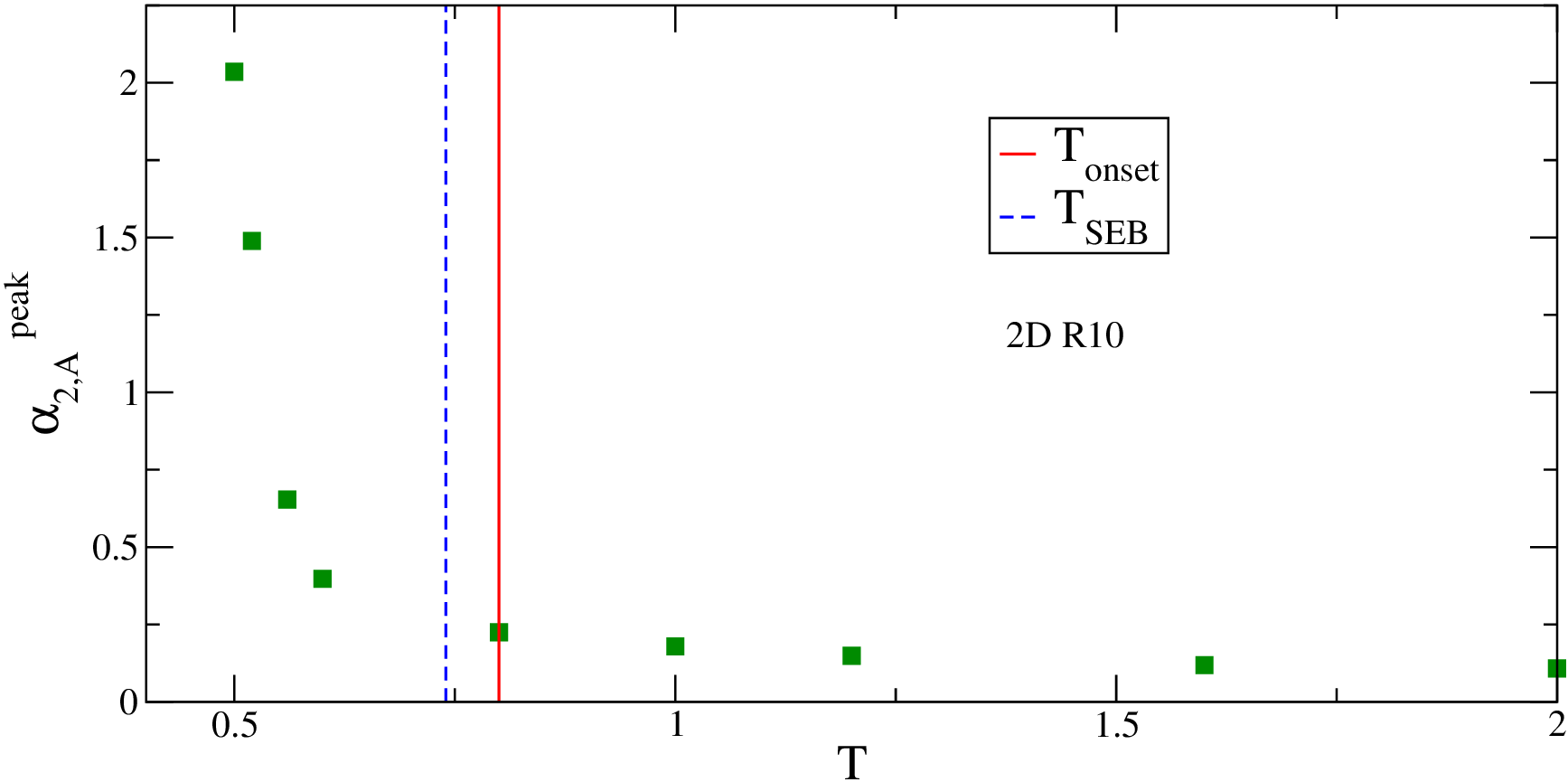} 
\caption{{\emph Left}: $T$ dependences of the peak height $\chi_{4}^{peak}$ of the dynamical susceptibility $\chi_{4}(t)$  
in the 2D R10 model. {\emph Right}: $T$ dependences of the peak height $\alpha_{2,A}^{peak}$ of the non-Gaussian parameter 
$\alpha_{2,A}(t)$ in the 2D R10 model. }
\label{fig:alpha2-chi4-2DR10}
\end{center}

\begin{center}
\includegraphics[width=6.7cm,height=5.5cm]{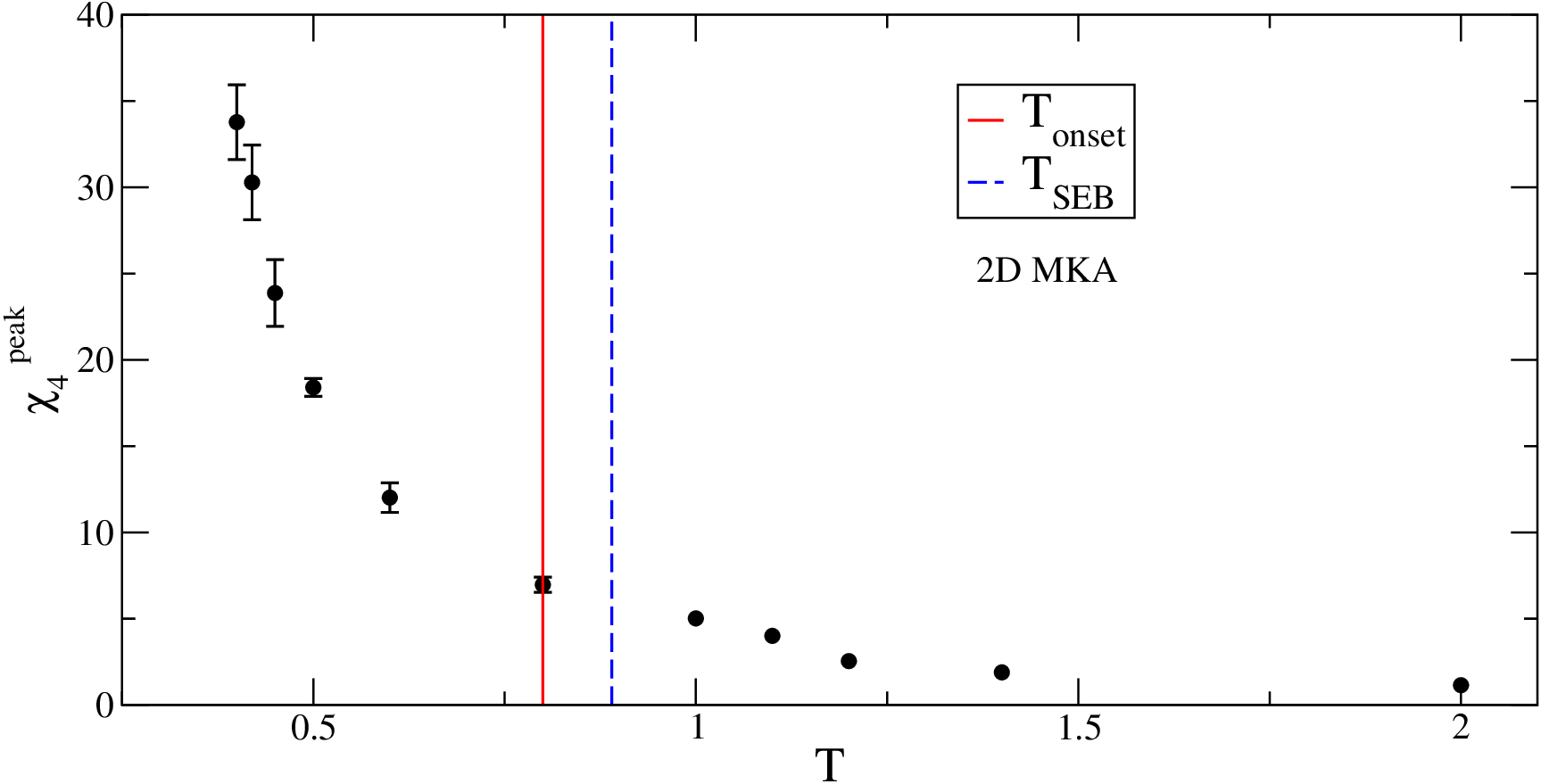} 
\hskip 0.5cm
\includegraphics[width=6.7cm,height=5.5cm]{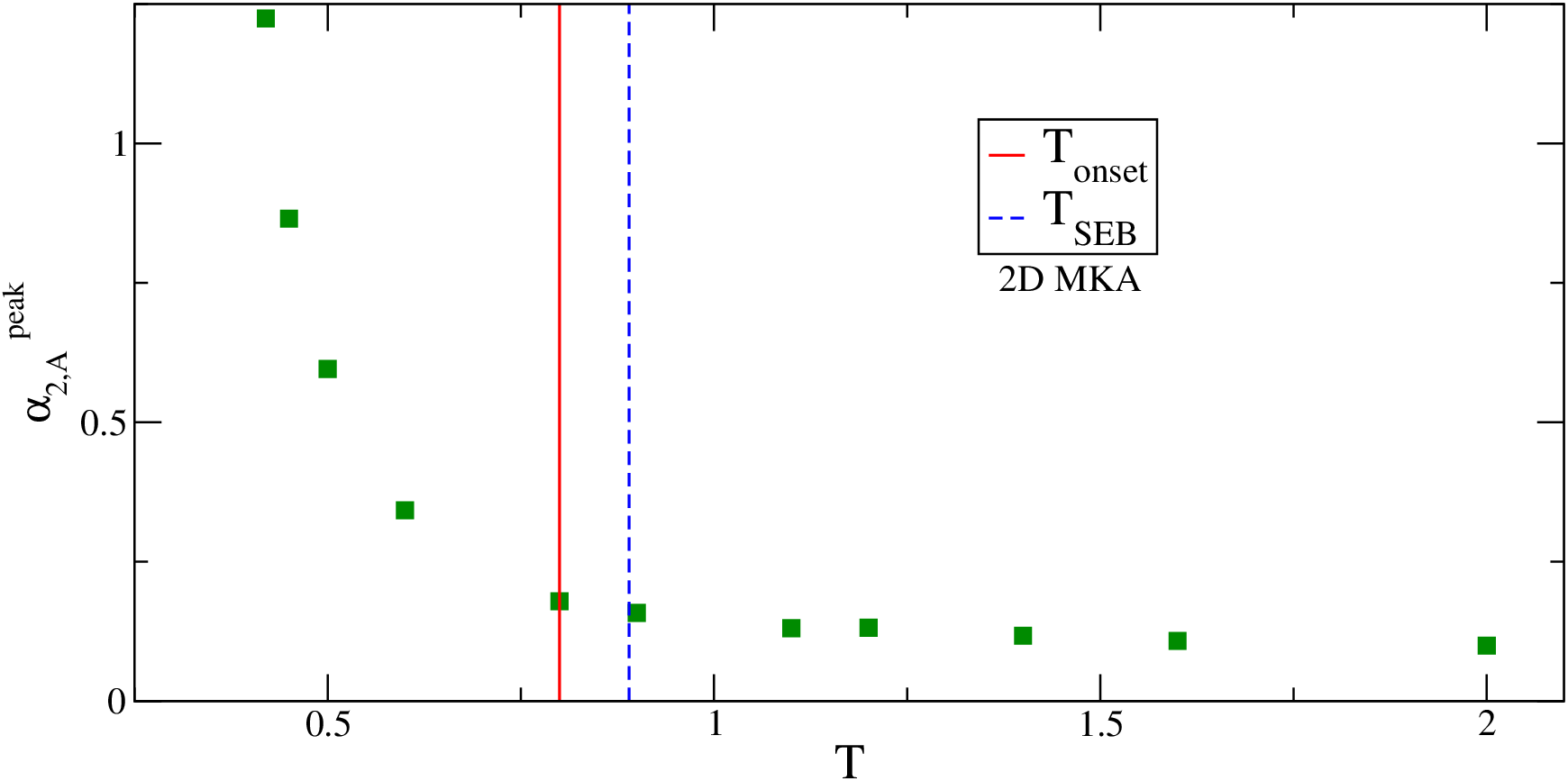} 
\caption{{\emph Left} : $T$ dependences of the peak height $\chi_{4}^{peak}$ of the dynamical susceptibility $\chi_{4}(t)$ in the 2DMKA model. 
{\emph Right} : $T$ dependences of  the peak height $\alpha_{2,A}^{peak}$ of the non-Gaussian parameter $\alpha_{2,A}(t)$ in the 2DMKA model. }\label{fig:alpha2-chi4-2DMKA}
\end{center}

\begin{center}
\includegraphics[width=6.7cm,height=5.5cm]{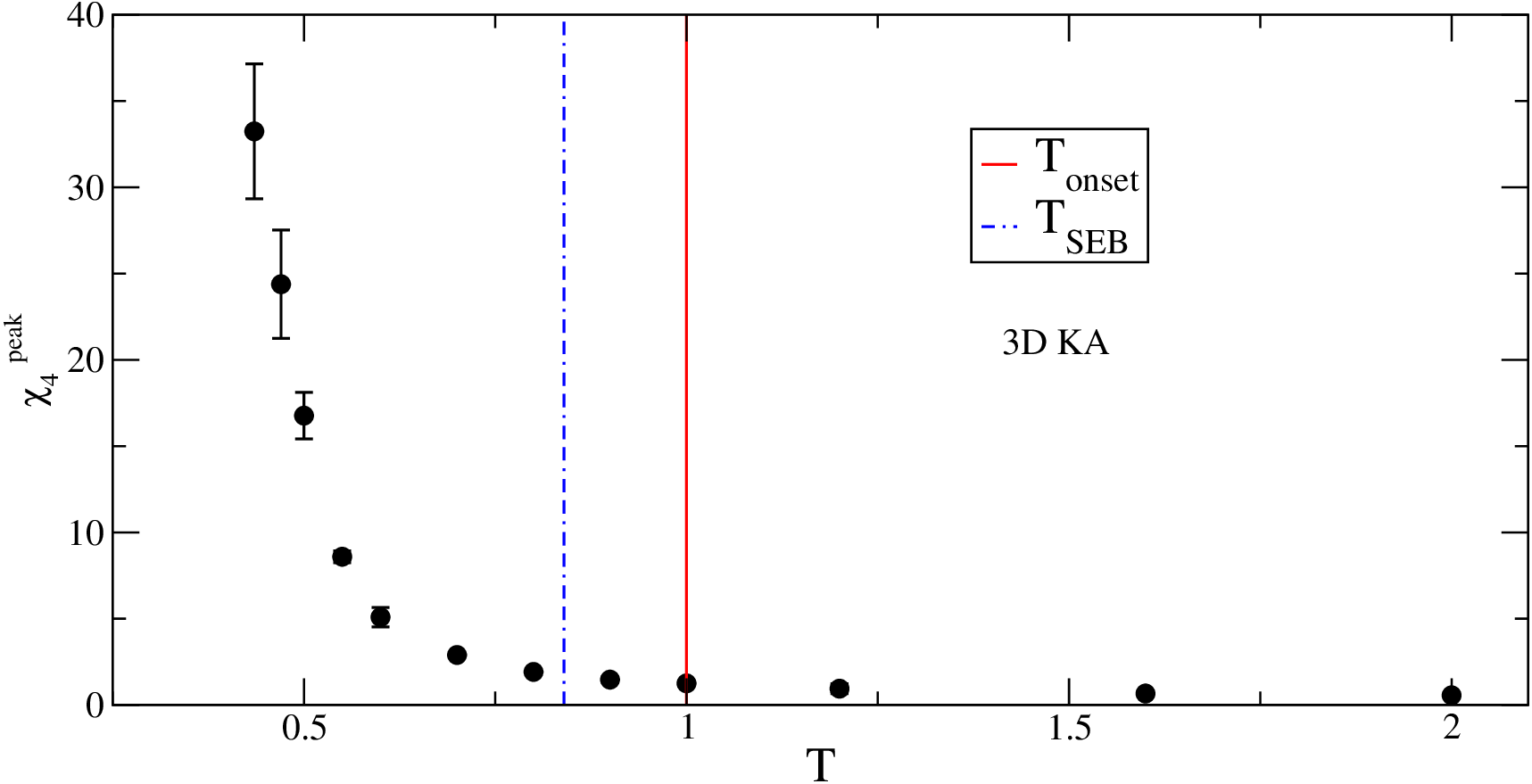} 
\hskip 0.5cm
\includegraphics[width=6.7cm,height=5.5cm]{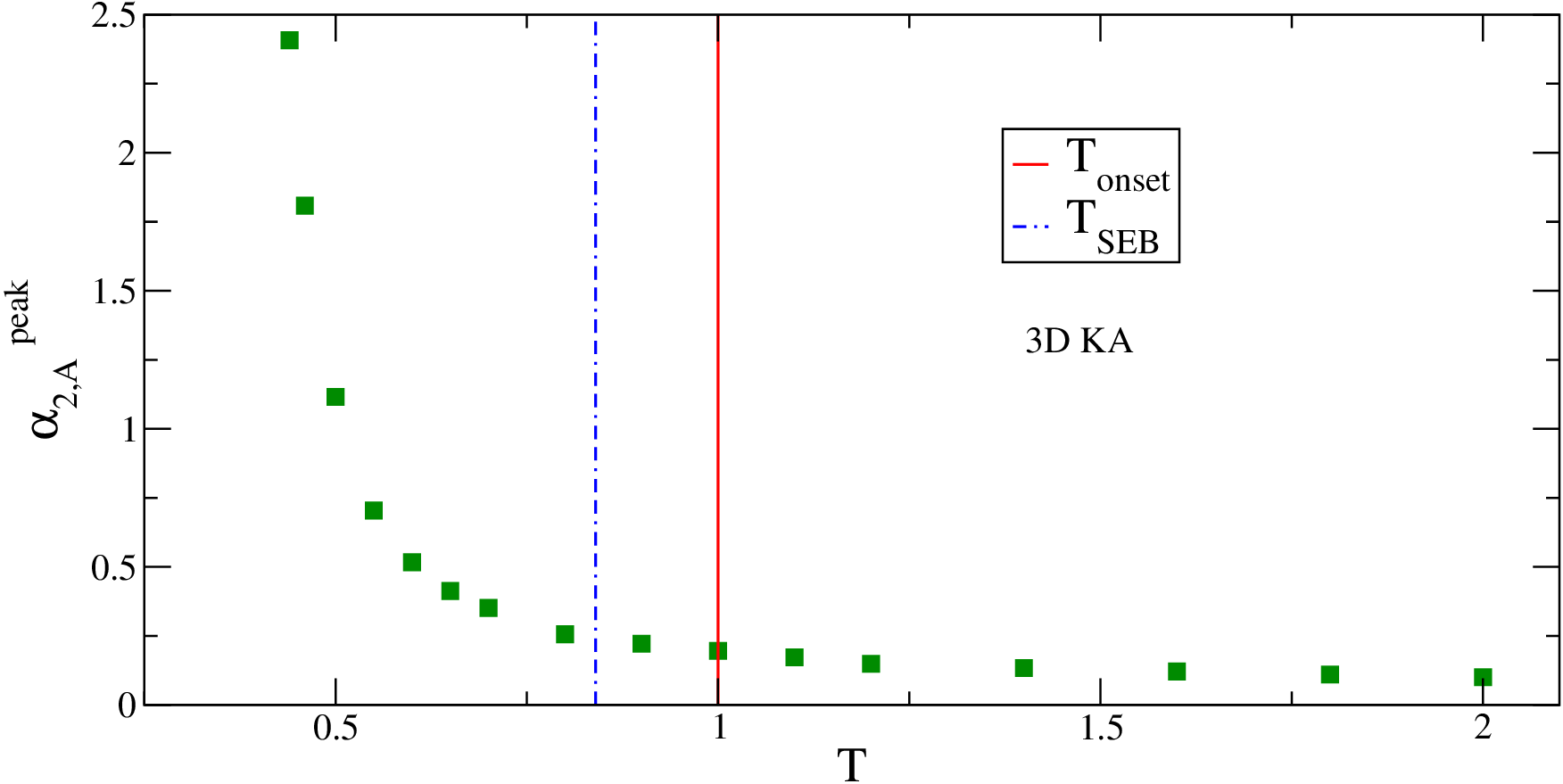} 
\caption{{\emph Left} : $T$ dependences of the peak height $\chi_{4}^{peak}$ of the dynamical susceptibility $\chi_{4}(t)$ in the 3DKA model. 
{\emph Right} : $T$ dependences of the peak height $\alpha_{2,A}^{peak}$ of the non-Gaussian parameter $\alpha_{2,A}(t)$ in the 3DKA model. }\label{fig:alpha2-chi4-3DKA}
\end{center}
\end{figure*}

\begin{figure*}[h!]
\begin{center}
\includegraphics[width=6.7cm,height=5.5cm]{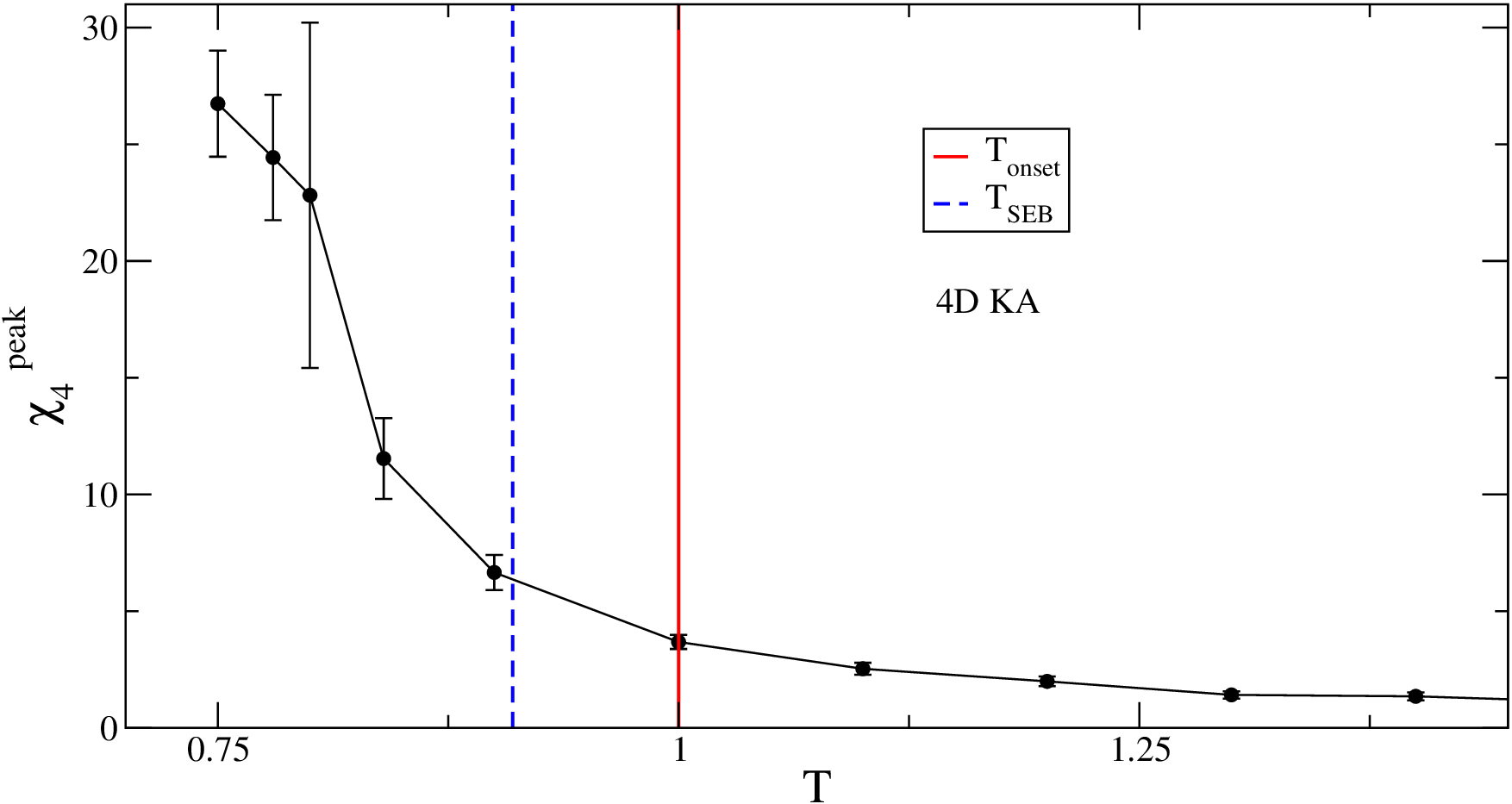} 
\hskip 0.5cm
\includegraphics[width=6.7cm,height=5.5cm]{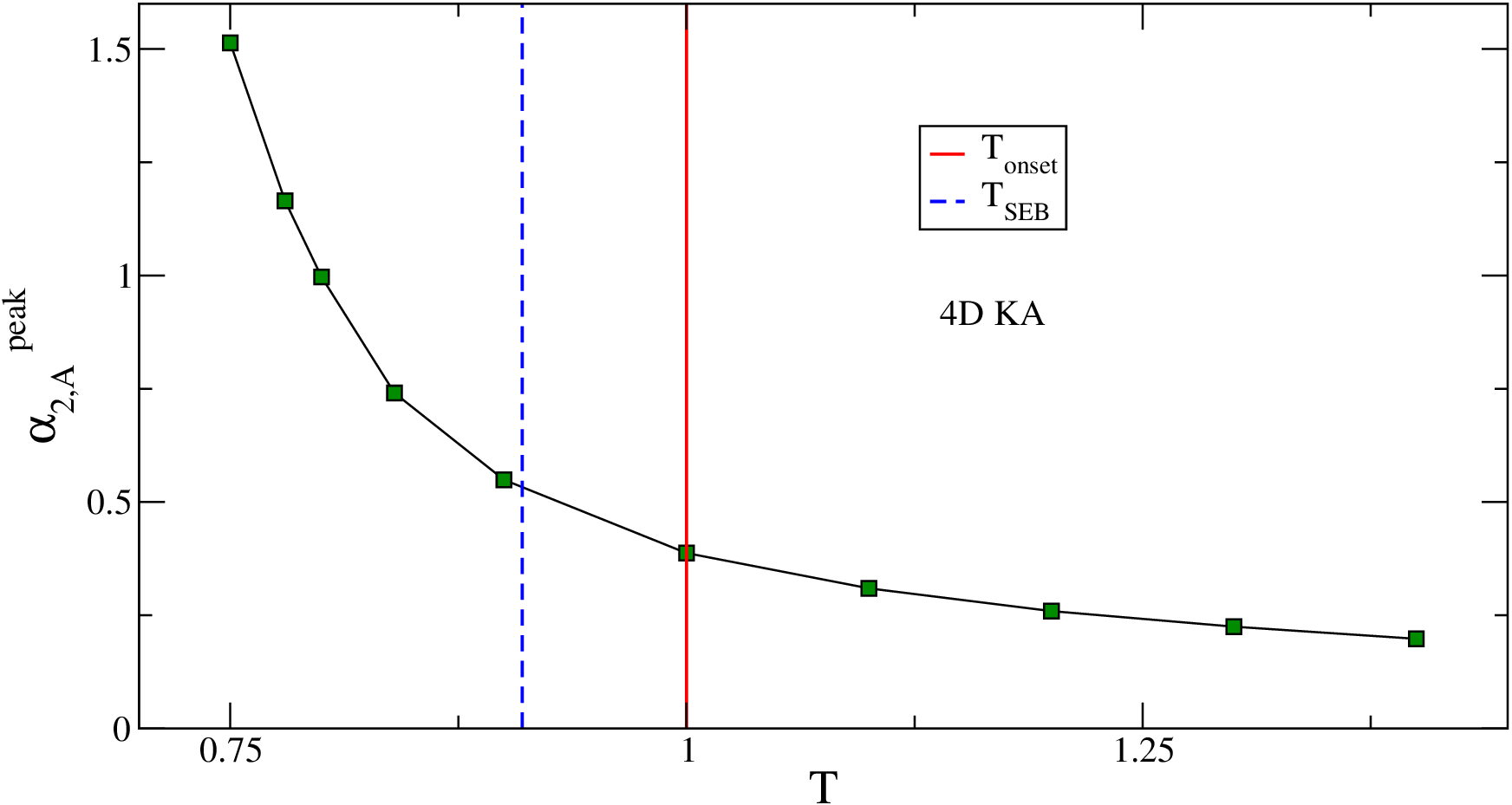} 
\caption{{\emph Left} : $T$ dependences of the peak height $\chi_{4}^{peak}$ of the dynamical susceptibility $\chi_{4}(t)$ 
in the 4DKA model. 
{\emph Right} : $T$ dependences of the peak height $\alpha_{2,A}^{peak}$ of the non-Gaussian parameter $\alpha_{2,A}(t)$ in the 4DKA model. }\label{fig:alpha2-chi4-4DKA}
\end{center}
\end{figure*}

\begin{figure*}[h!]
\begin{center}
\includegraphics[width=6.7cm,height=6cm]{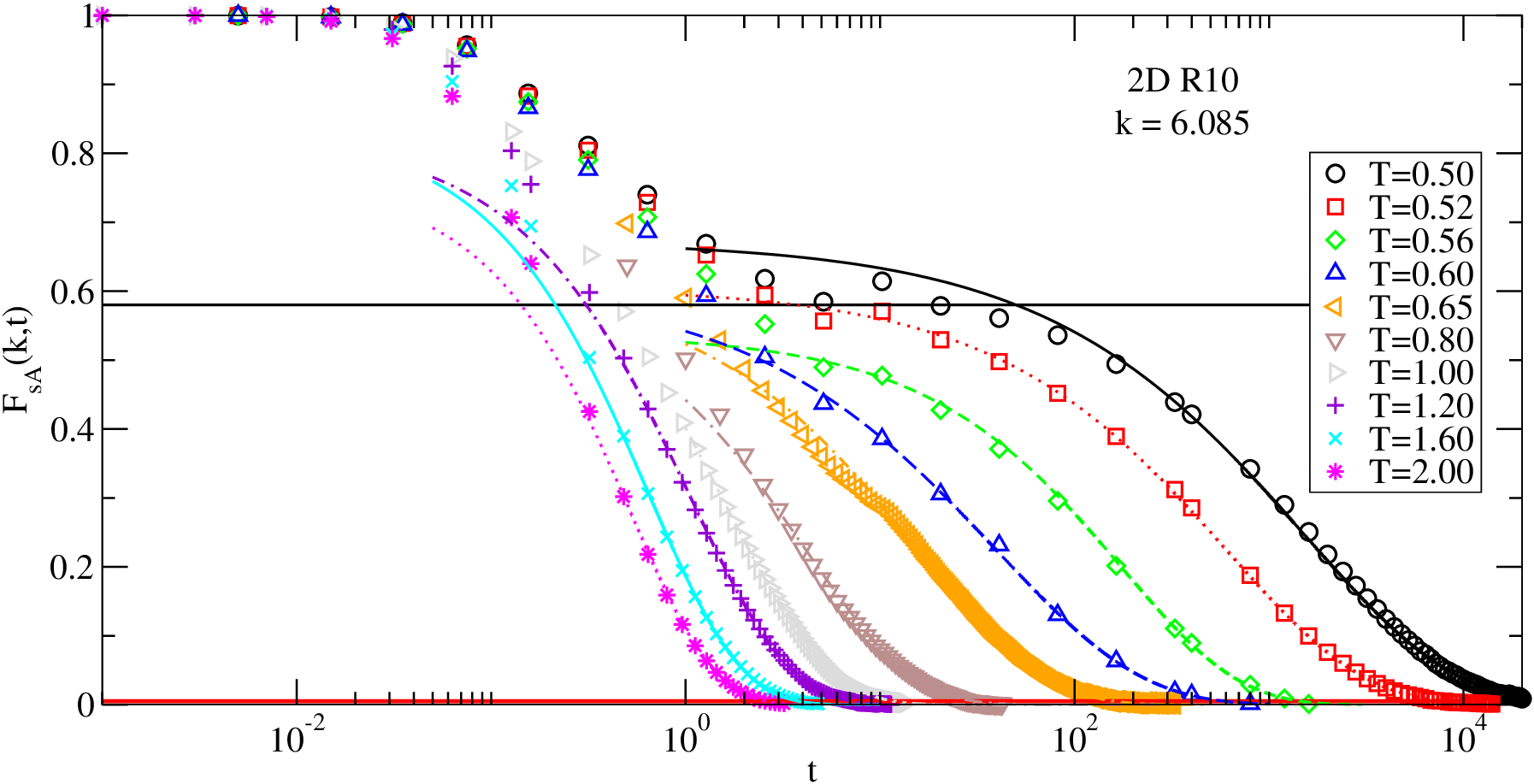} 
 \hskip 0.5cm
\includegraphics[width=6.7cm,height=6cm]{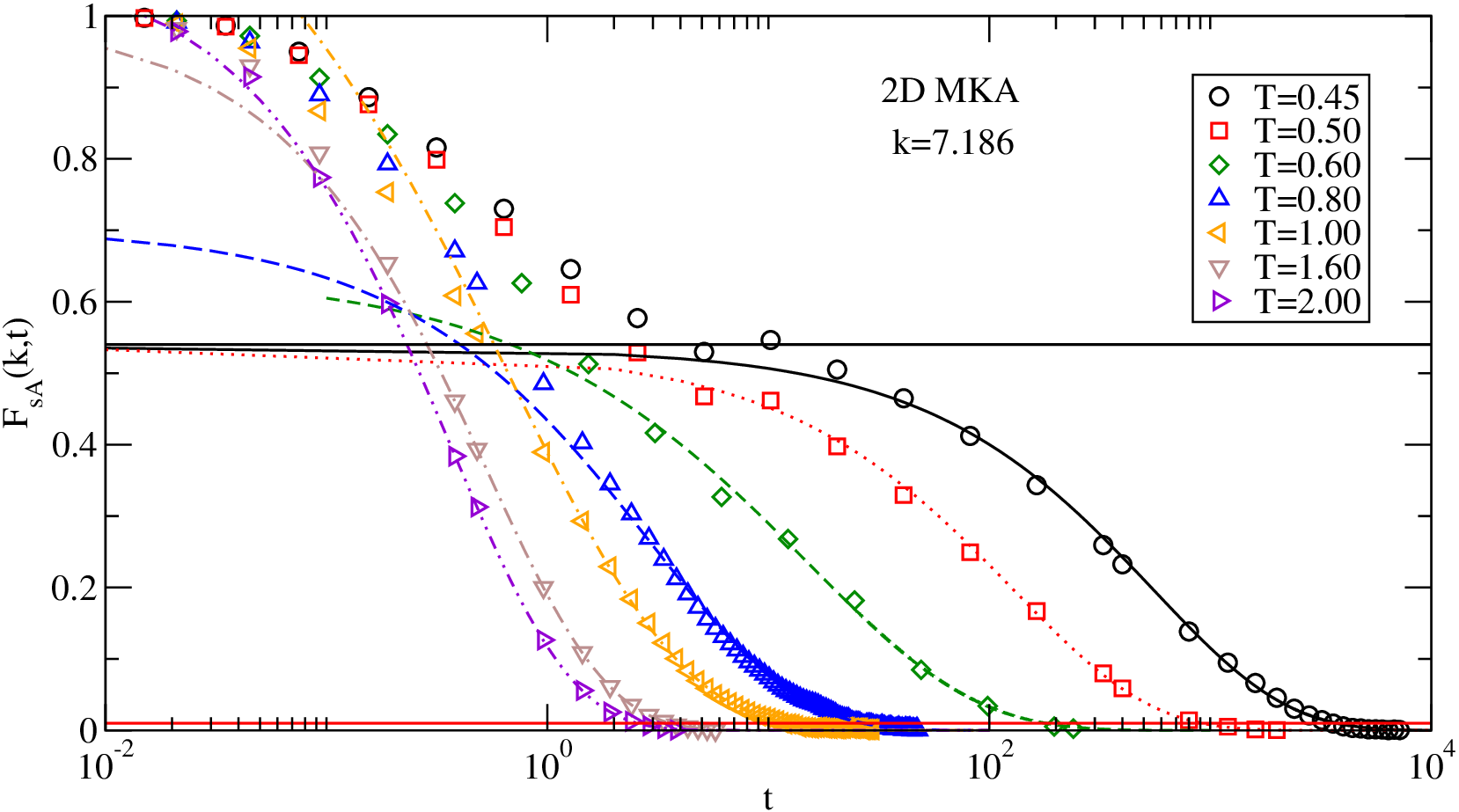} 
\vspace{5mm}
\includegraphics[width=6.7cm,height=6cm]{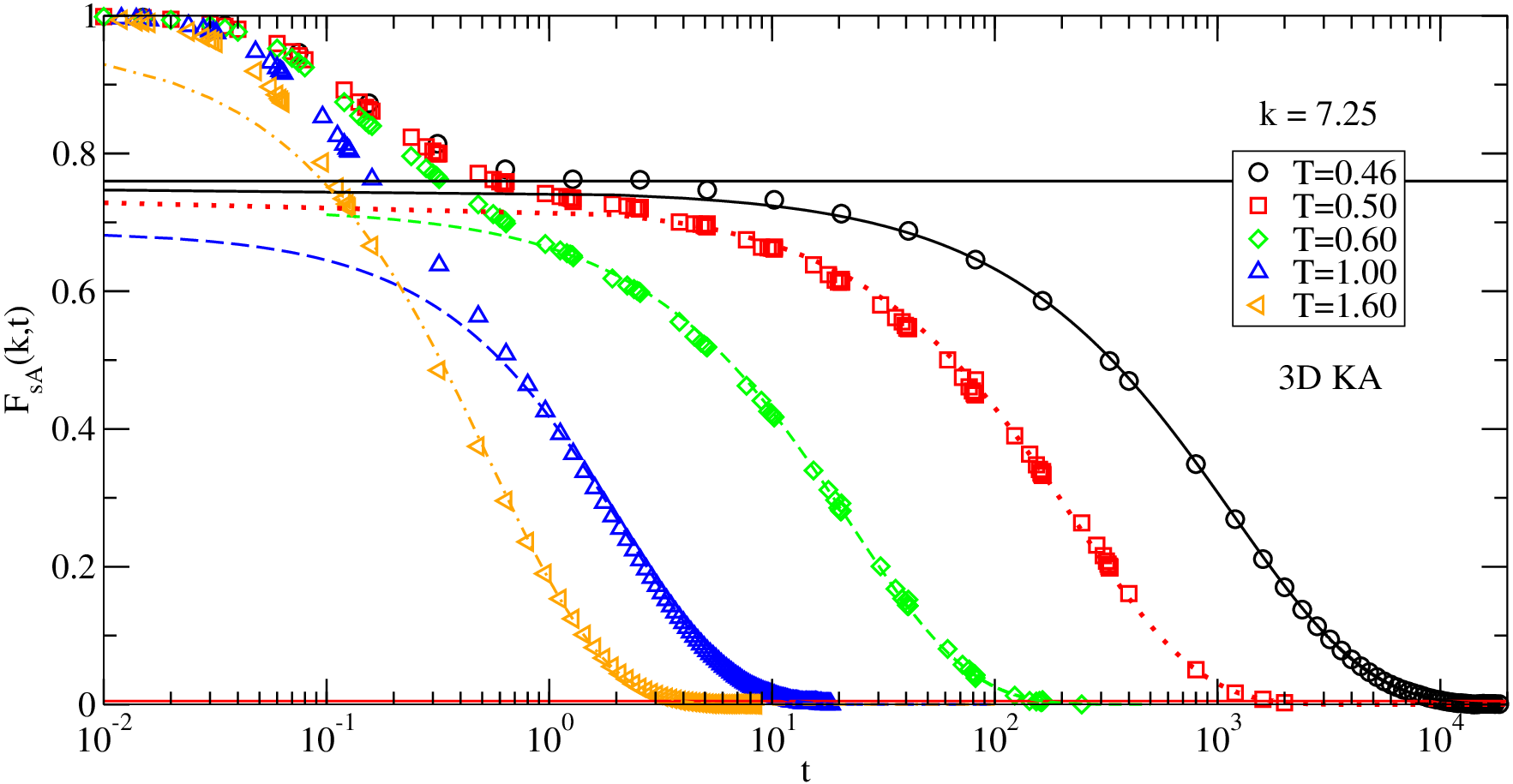} 
\hskip 0.5cm
\includegraphics[width=6.7cm,height=6cm]{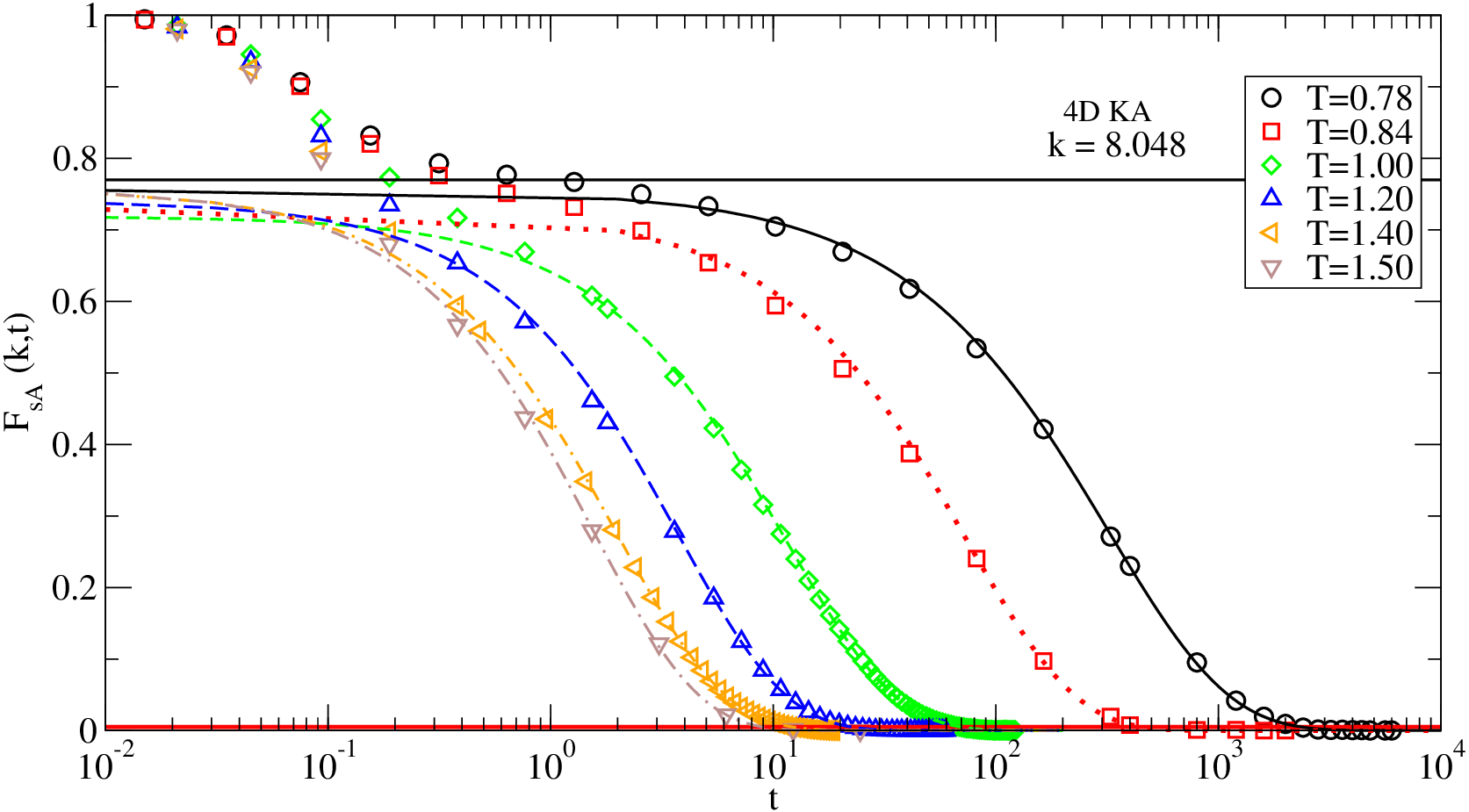} 
\caption{$T$ dependence of $F_{sAkt}$ in 2DR10, 2DMKA, 3DKA and 4DKA models. Also shown are fit curves to the stretched exponential form. Horizontal lines indicate the range of $y$ values within which the fitting is performed.}
\label{fig:FsAkt-diff-dim}
\end{center}
\end{figure*}

\clearpage
\subsection{Dynamical heterogeneity and the breakdown of the Stokes-Einstein relation}
Here we consider three measures of dynamical heterogeneity, namely the
dynamical susceptibility $\chi_{4}$, the non-Gaussian parameter
$\alpha_2$ and the stretching exponent $\beta_{KWW}$. We examine to
what extent these measures of heterogeneity correlate with the degree
of the breakdown of the SE relation. 

Both $\chi_{4}$ and $\alpha_2$ have been extensively studied as
indicators of dynamical heterogeneity ({\it e. g.}
\cite{pap:Ovlap-Glotzer-etal,pap:Ovlap-Donati-etal,pap:Karmakar-PNAS,pap:Starr-etal}. These
quantities are functions of time, and exhibit a maximum value at a
characteristic time that is close to $\tau_{\alpha}$
\cite{pap:Karmakar-PNAS} in the case of $\chi_{4}$ and to $t^{*} \sim
(D/T)^{-1}$ in the case of $\alpha_2$ \cite{pap:Starr-etal}. The time dependence of 
$\chi_{4}$ and $\alpha_2$ are shown in Fig. \ref{fig:alpha2-chi4-vst-4DKA}. Peak values 
$\chi_{4}^{peak}$ and $\alpha\_{2}^{peak}$ extracted from such time dependence are described 
further below as measures of heterogeneity.

The temperature dependence of the peak of the dynamical susceptibility
$\chi_{4}^{peak}$ and that of the non-Gaussian parameter
$\alpha_{2}^{peak}$ in different models are shown in
Figs. \ref{fig:alpha2-chi4-2DR10}, \ref{fig:alpha2-chi4-2DMKA},
\ref{fig:alpha2-chi4-3DKA} and \ref{fig:alpha2-chi4-4DKA}. Clearly, in
\emph{all dimensions}, the peak heights $\chi_{4}^{peak}$ and
$\alpha_{2}^{peak}$ grows as the temperature decreases thus indicating that
the dynamics is heterogeneous in the lower part of the temperature range that we study. 

As a third direct measure of the degree of dynamical heterogeneity, we
estimate the stretching exponent $\beta_{KWW}$. We have computed
$\beta_{KWW}$ from the self intermediate scattering function
$F_{sAkt}$ for $k$ at the peak of the structure factor. Fig.
\ref{fig:FsAkt-diff-dim} shows the temperature dependence of $F_{sAkt}
(k,t)$ for different models and the fits to stretched exponential form
from which the stretching exponents $\beta_{KWW}$ are
obtained \cite{footnote}.

\begin{figure}[h!]
\begin{center}
\includegraphics[width=8cm,height=6cm]{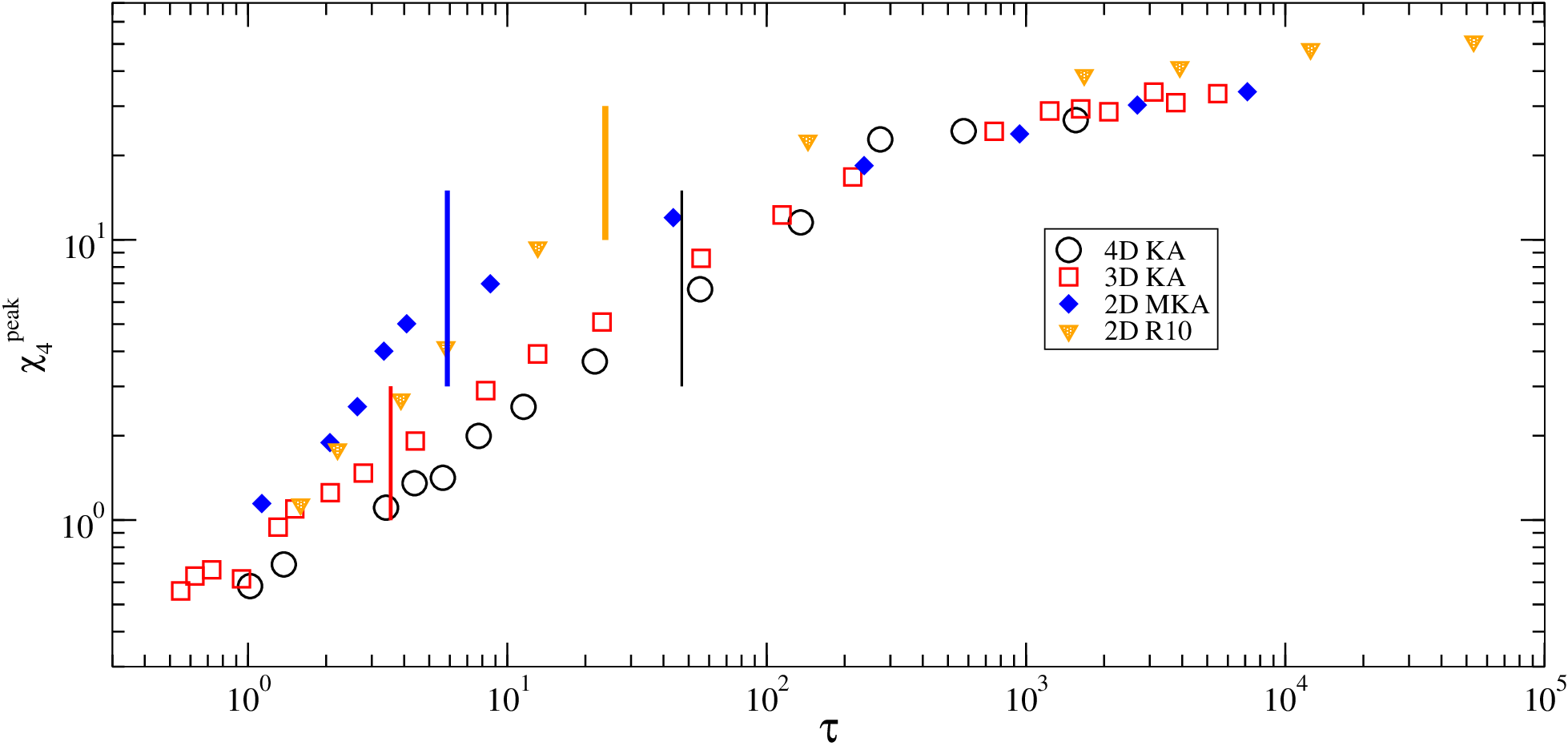} 
\caption{Comparison of the \emph{degree of heterogeneity in different dimensions} using the $\chi_{4}^{peak}$ as a measure. Vertical lines correspond to 
$T_{SEB}$}
\label{fig:degree-heterogen-chi4}
\end{center}
\end{figure}

In order to compare the degree of the SE breakdown with that of
dynamical heterogeneity, we plot $\chi_{4}^{peak}$,
$\alpha_{2,A}^{peak}$ and $\beta_{KWW}$ for the different models
against the relaxation time $\tau_{\alpha}$. We expect that the system
showing a stronger SE breakdown at a given temperature (or relaxation
time) will also show a larger degree of heterogeneity. The data in
Fig. \ref{fig:degree-heterogen-chi4} for $\chi_{4}^{peak}$ shows that
indeed, the four dimensional system by and large has smaller values of
$\chi_{4}^{peak}$ than the three dimensional system, although the
values are close especially at low temperatures. This is consistent
with the larger SE breakdown observed in three dimensions. Two
dimensional systems show substantially larger values of
$\chi_{4}^{peak}$ than three dimensional systems for most of the
studies points. While this is qualitatively consistent in principle
with the marginal increase in the degree of SE breakdown in two
dimensions, quantitatively the difference in $\chi_{4}^{peak}$ is
large and surprising. 

In similar fashion, $\alpha_{2,A}^{peak}$ values show an increase in
going from four to three dimensions. However, two dimensional systems
show smaller values of $\alpha_{2,A}^{peak}$ than either three or four
dimensional systems. This is contrary to the relative degrees of SE
breakdown in these different systems. Such an inconsistency may reveal
either a shortcoming of $\alpha_{2,A}^{peak}$ as a measure of
heterogeneity, of another aspect of the peculiar behavior of two
dimensional systems.  Further investigations are necessary to
understand this behavior. 

Finally, we find that the stretching exponents $\beta_{KWW}$ depend
systematically on the dimensionality, even though the estiamtes are
noisy. The $\beta_{KWW}$ values for the four dimensional system (4DKA)
lie above those of the three dimensional system (3DKA), which in turn
are larger than those the values for the two dimensional systems
(2DMKA and 2DR10). Though not strictly consistent with the degree of
SE breakdown as quantified by the breakdown exponent, the behavior of
$\beta_{KWW}$ does not throw up any surprises. A more detailed study,
considering also length scales over which dynamics is correlated
spatially, perhaps along the lines of \cite{chong}.

\begin{figure}[h!]
\begin{center}
\includegraphics[width=8cm,height=7cm]{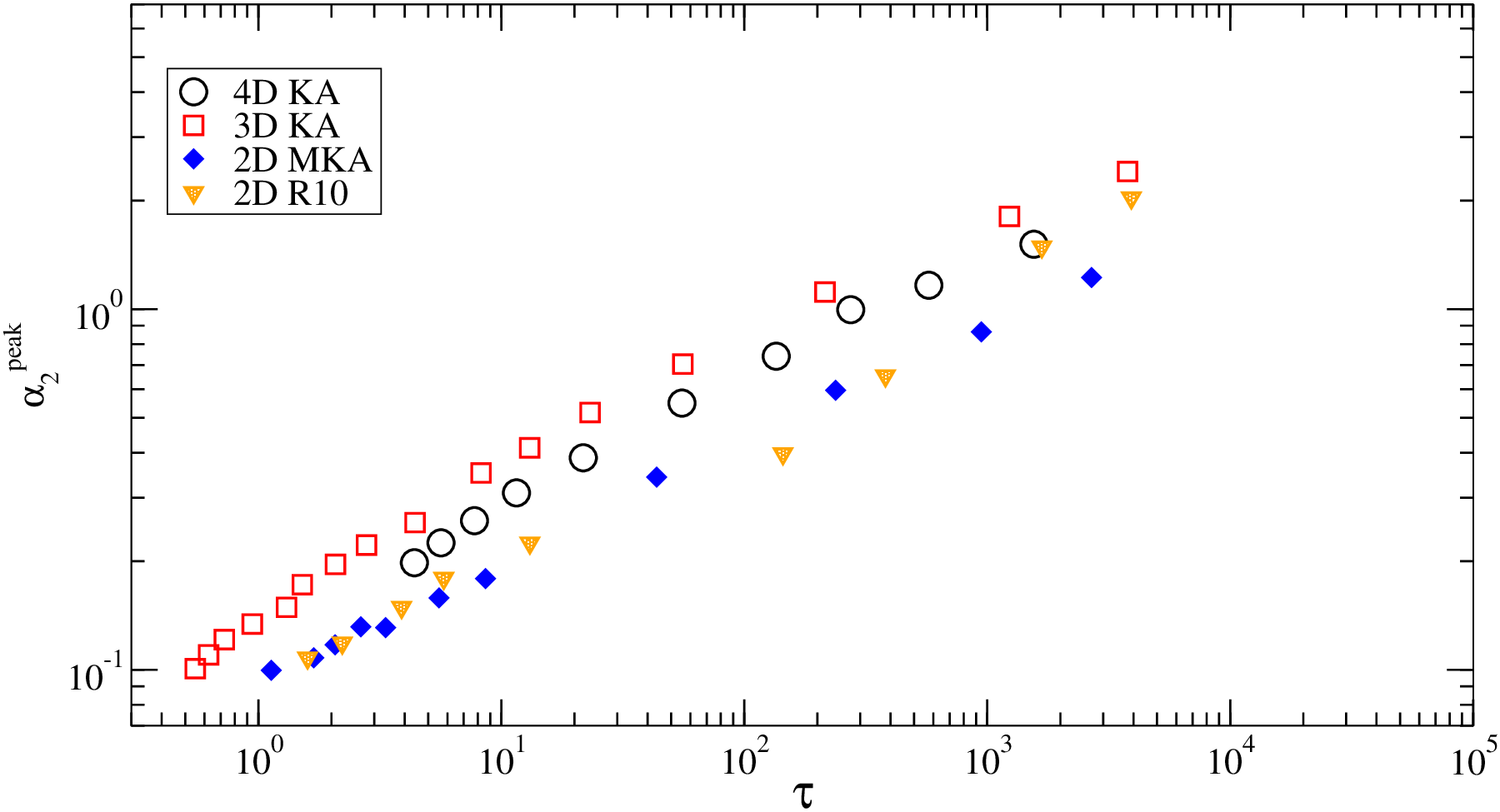} 
\caption{Comparison of the \emph{degree of heterogeneity in different dimensions} using  $\alpha_{2,A}^{peak}$ as a measure.} 
\label{fig:degree-heterogen-SEratio}
\end{center}
\end{figure}

\begin{figure}[h!]
\begin{center}
\includegraphics[width=8cm,height=6cm]{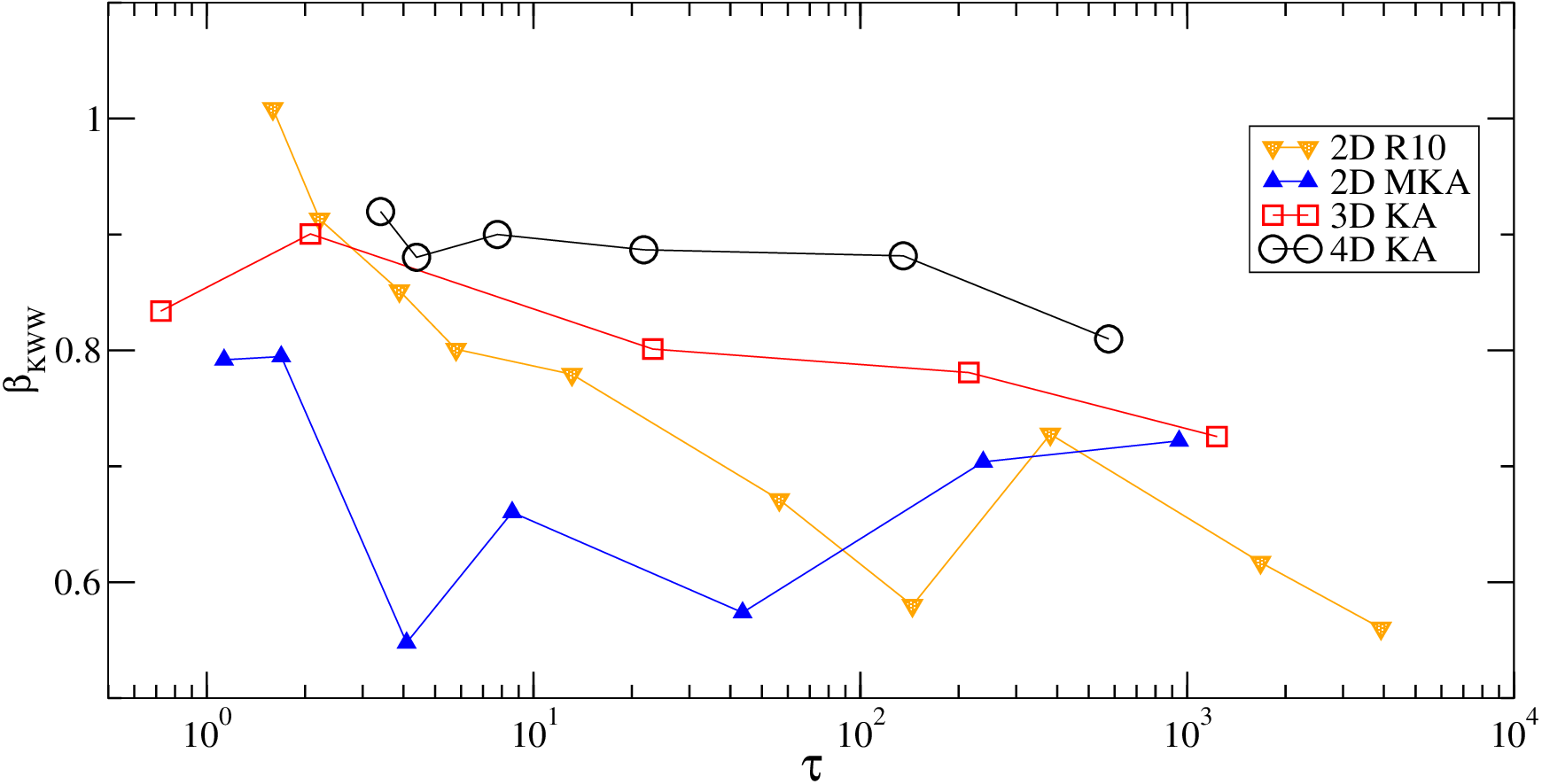} 
\caption{Comparison of the \emph{degree of heterogeneity in different dimensions} using the $\beta_{KWW}$  as a measure.}
\label{fig:betaKWW-alldim}
\end{center}
\end{figure}

\begin{figure}[h!]
\begin{center}
\includegraphics[height=6cm,width=8cm]{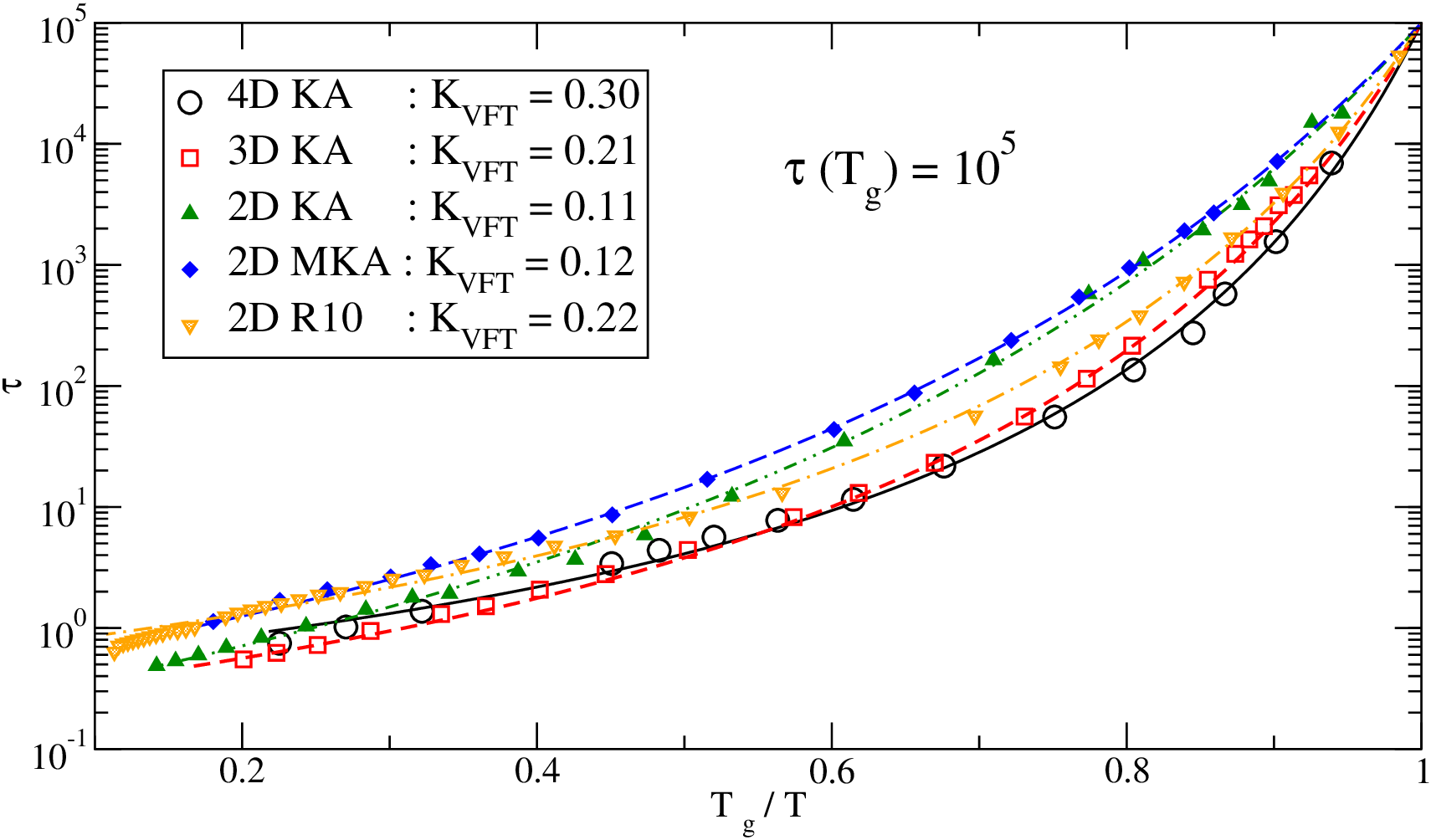} 
\vspace{8mm}
\includegraphics[height=6cm,width=8cm]{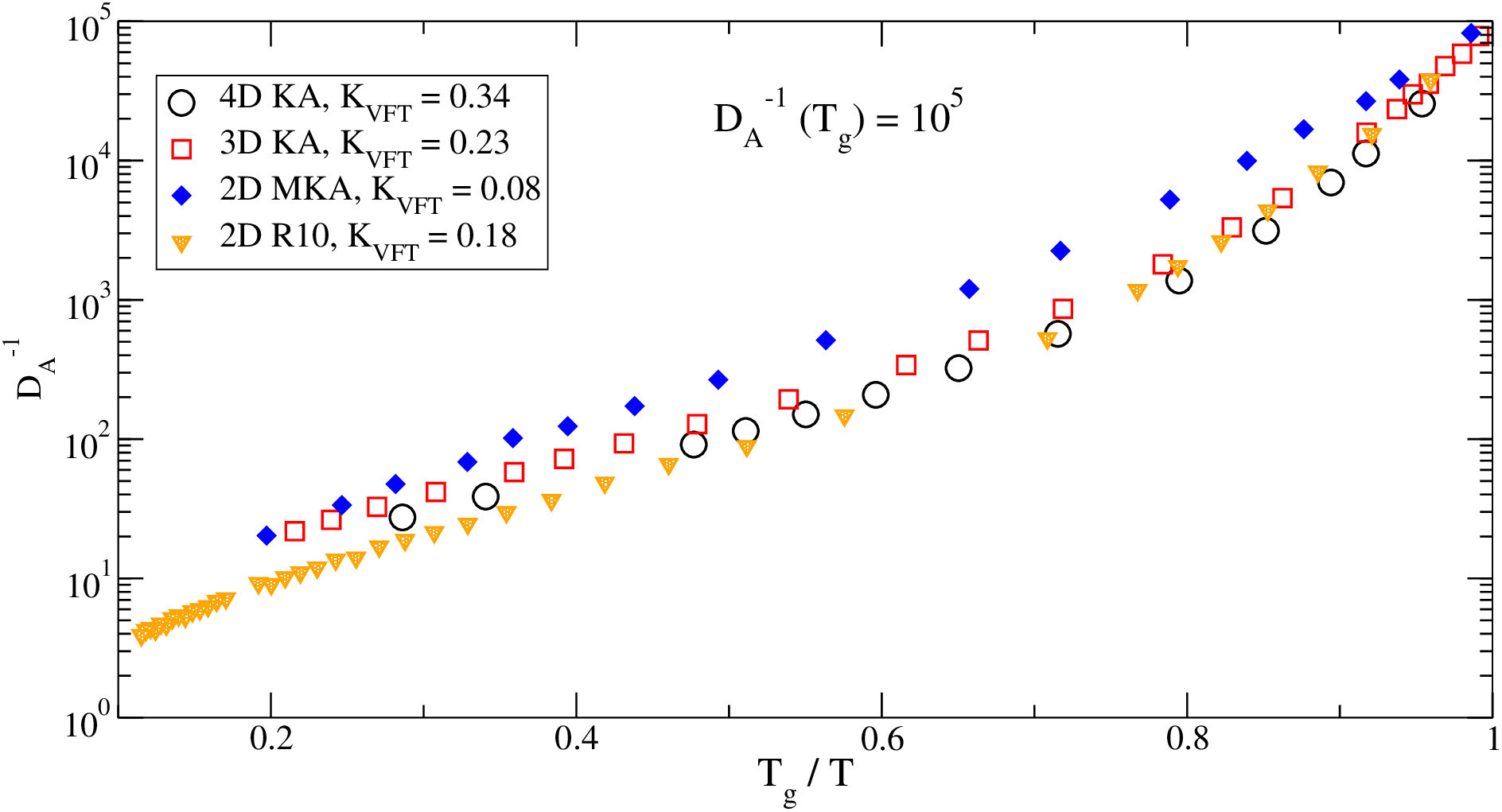} 
\caption{\emph{Top:} Fragility plot of relaxation times for five
  models in 2,3,4 spatial dimensions.  VFT fits to relaxation times
  shown are used to obtain the kinetic fragility $K_{VFT}$ (see
  discussion). The fragility plot employs a ``simulation glass
  transition temperature'' $T_{g}$ defined as $\tau(T_{g})=10^{5}$
  (reduced unit) to scale temperatures. The $K_{VFT}$ values are
  listed in Table \ref{tab:parameters}.  \emph{Bottom:} Fragility plot
  for four models in 2,3,4 spatial dimensions using the inverse of the
  diffusion coefficient.  \emph{The plots show that systems at higher
    dimensions are more fragile.}}
\label{fig:VFT-vs-dim}
\end{center}
\end{figure}

\subsection{Dependence of the fragility on spatial dimensions}
As discussed in the introduction, the fragility of a glass former has
been argued to be correlated with the heterogeneity of a glass forming
liquid, and by extension the degree of the breakdown of the SE
relation. To evaluate such an expectation, we calculate the fragility
of the model glass formers studied. We estimate the kinetic fragility
from the Vogel-Fulcher-Tammann (VFT) fit to the $T$-dependence of the
$\alpha$ relaxation times:

\begin{equation}
\tau(T) = \tau_{0} \exp\left[\frac{1}{K_{VFT}(\frac{T}{T_{VFT}}-1)} \right] \label{eqn:kinfr2}
\end{equation}

\begin{figure}[h!]
\begin{center}
\includegraphics[width=6.5cm,height=6cm]{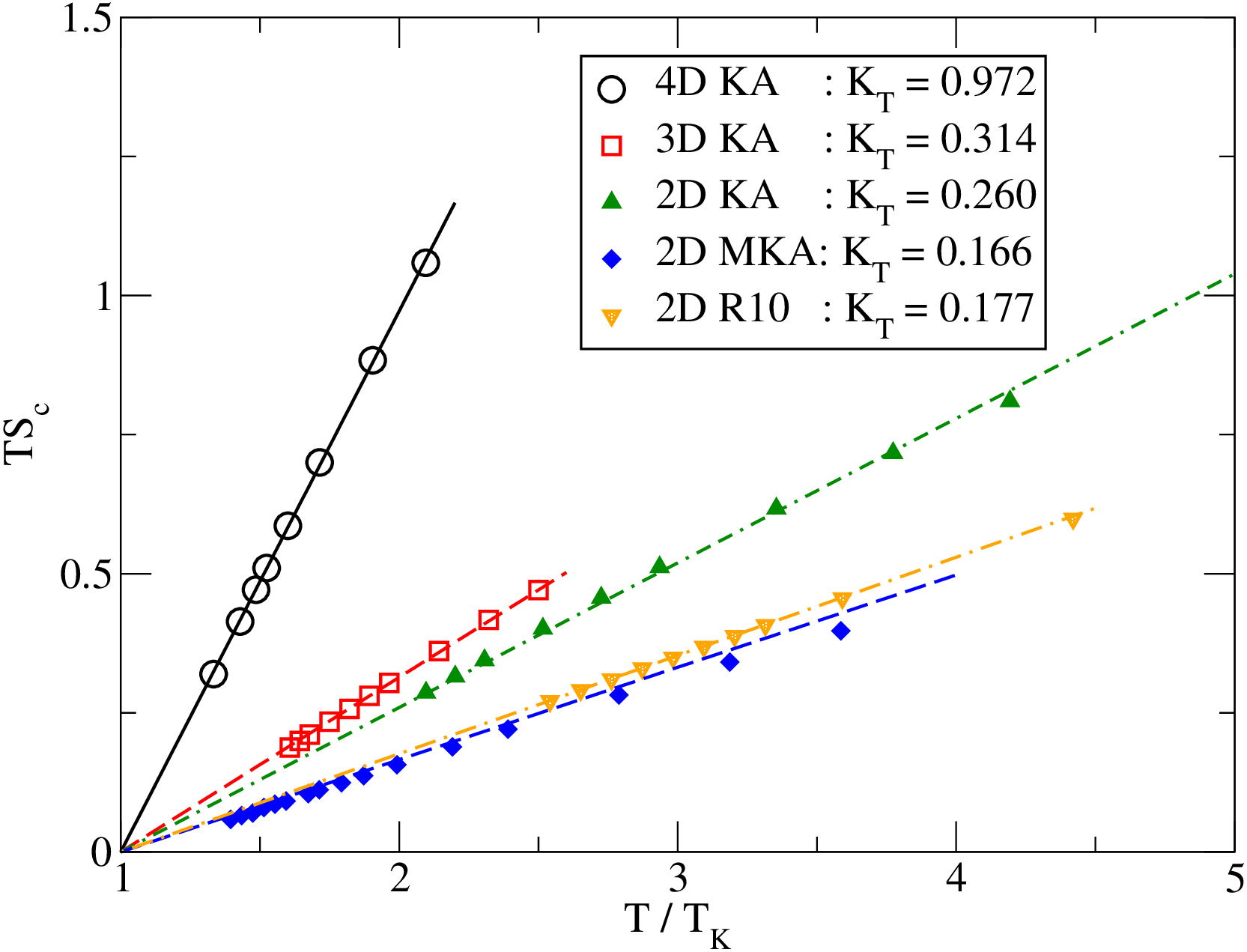} 
\caption{$T$ dependence of $TS_{c}$ of five models in 2,3,4 spatial dimensions plotted as $TS_{c}$ 
{\it vs.} $T/ T_{K}$ so that the slope is an estimate of the thermodynamic fragility $K_{T}$ 
(listed in Table \ref{tab:parameters}). The plot shows that the thermodynamic 
fragility increases with increasing spatial dimensionality.}
\label{fig:TSc-vs-dim}
\end{center}
\end{figure}

\begin{figure}[h!]
\begin{center}
\includegraphics[width=8cm,height=7cm]{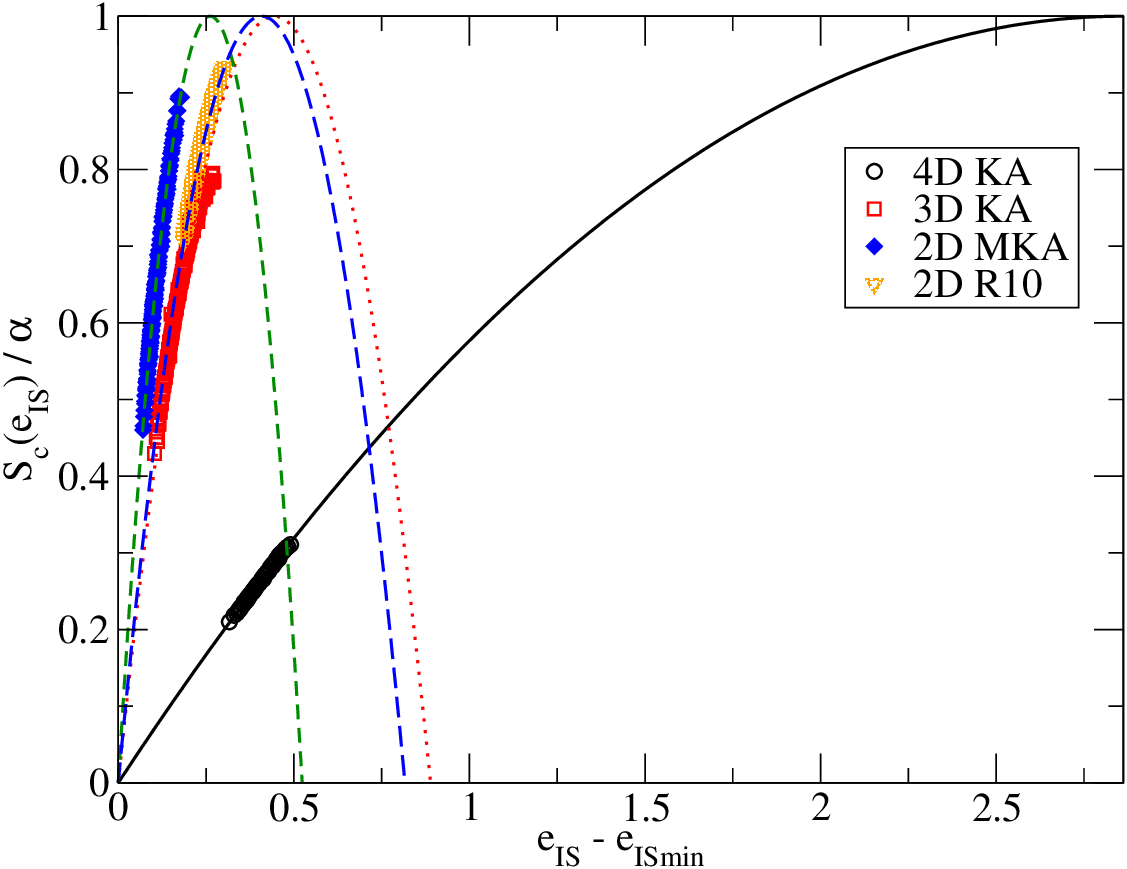} 
\caption{The configurational entropy density $S_{c}(e_{IS})$ of the inherent structures plotted 
{\it vs.} their energy shifted by its minimum possible value $e_{ISmin}$ defined from $S_{c}(e_{ISmin}) =0$. 
The lines are fits to the parabolic form: $\frac{S_{c}(e_{IS})}{\alpha} = 
1 - \frac{(e_{IS}- e_{IS}^{0})^{2}}{(\sigma \sqrt{\alpha})^{2}}$ where $e_{IS}^{0} = e_{ISmin} + \sigma \sqrt{\alpha}$ is the 
IS energy at the peak of the distribution. The distribution is broader for higher dimensions which partially explains 
the increase of the thermodynamic fragility with increasing spatial dimension.}
\label{fig:SceIS-vs-dim}
\end{center}
\end{figure}

Fig. \ref{fig:VFT-vs-dim} shows the $T$ dependence of the $\alpha$
relaxation times in the five models we have studied, in a ``fragility
plot'' which shows relaxation times in an Arrhenius plot with the
temperature scaled by the glass transition temperature. We use a
``simulation glass transition temperature'' $T_g$ defined from
$\tau(T_{g})=10^{5}$ (reduced units) in order to make such a plot. The
bottom panel shows a fragility plot for the inverse diffusion
coefficients, with $T_g$ defined from $D_{A}^{-1} = 10^{5}$ (reduced
units).  The plots and the VFT fits to the relaxation times clearly
show that the fragility of the liquids is larger at higher
dimensions. This trend is opposite to the expectation based on that of
the breakdown of the SE relation, and an assumption that the two are
correlated. Nevertheless, there are no strong reasons to expect that
fragility is correlated with heterogeneous dynamics, in particular
with a variation of dimensionality. For example, the analysis by Xia 
and Wolynes \cite{pap:Frag-Xia-Wolynes} is restricted to three dimensions, 
and it would be interesting to carry it out for other dimensions as well.

On the other hand, a rationalization of the fragilities themselves of
the different systems studied can be attempted using the energy
landscape approach in \cite{pap:AG-Sastry}, based on the validity of
the Adam-Gibbs relation. In order for the VFT relation and the Adam-Gibbs
relation to both hold, one needs the $T$ dependence of
$TS_{c}$ to be linear:

\begin{equation}
TS_{c} = K_{T}\left(\frac{T}{T_{K}} - 1\right), \label{eqn:pelfrag}
\end{equation}
Eq. \ref{eqn:pelfrag} defines the thermodynamic fragility $K_{T}$. 
The thermodynamic fragility $K_{T}$ can be related to the kinetic fragility 
$K_{VFT}$ by combining the AG relation $\tau=\tau_{0}\exp(\frac{A}{TS_{c}})$, 
the linear $T$ dependence of $TS_{c}$ (Eq. \ref{eqn:pelfrag}) and 
the VFT form (Eq. \ref{eqn:kinfr2}) to obtain \cite{pap:Frag-Sengupta}, 

\begin{equation}
K_{VFT} = K_{T} / A 
\label{eqn:KVFT-KT}
\end{equation}

The origin of $K_{T}$ can be understood from the properties 
of the minima of the potential energy landscape \cite{pap:AG-Sastry}, 
known as ``inherent structures'' (IS). The configurational entropy is 
the entropy associated with the multiplicity of energy minima, and if 
the density of states of such minima with respect to their energy 
$e_{IS}$ is described by 

\begin{equation}
S_{c}(e_{IS})/N k_B = \alpha - {(e_{IS} - e_{IS}^{0})^2 \over \sigma^2}
\label{eqn:ScDOS}
\end{equation}
such that $\alpha$ quantifies the total number of inherent structures, 
and $\sigma^2$ the variance of the distribution, the thermodynamic 
fragility (under the simplifying assumption of the vibrational entropies 
of all the inherent structures being the same) is given by  

\begin{equation}
K_{T} = \sqrt{\alpha} \sigma/ 2.
\label{eqn:ScDOS}
\end{equation}

In other words, $K_{T}$ is proportional to the total spread of the density of states. 
Fig. \ref{fig:TSc-vs-dim} shows the $T$ dependence of the
configurational entropy for five models. We see that the
thermodynamic fragility increases significantly as the spatial
dimension increases. Further, in Fig. \ref{fig:SceIS-vs-dim} we show
that the quantity $\sigma \sqrt{\alpha}$ estimating the width of the
configurational entropy density increases with increasing spatial
dimension, which partially explains the increase of the thermodynamic
fragility with increasing spatial dimensions.

Eq. \ref{eqn:KVFT-KT} formally resolves the contributions of
configurational entropy ($K_{T}$) and the energy barrier ($A$) to the
kinetic fragility. Figs. \ref{fig:TSc-vs-dim} and
\ref{fig:SceIS-vs-dim} shows that the configurational entropy
contribution increases with increasing dimension thus by itself 
explaining the increase in the kinetic fragility at higher
dimensions. The characteristic parameters related to fragility for the
five models studied here are summarized in Table \ref{tab:parameters}.

\begin{table*}[h!]
\begin{center}
\caption{Fragility-related parameters for the models in different dimensions. $T_{K}$ 
is the Kauzmann temperature obtained from extrapolating $TS_{c}$ to zero. $K_{VFT}$ 
is the kinetic fragility obtained from VFT fits to relaxation times. $K_{T}$ is the 
thermodynamic fragility obtained from the $T$-dependence of $TSc$. $A$ is the Adam-Gibbs coefficient. 
$K_{AG}=K_{T}/A$ is the kinetic fragility estimated from the AG relation.}
\begin{tabular}{|p{19mm}|p{13mm}|c|c|c|c|c|c|p{13mm}|}
\hline
Model&  Density & $T_{K}$ & $T_{VFT}$ & $T_{g}$ & $K_{VFT}$  & $K_{T}$ & $A$ & $K_{AG}=K_{T}/A$\\
\hline
4D KA  & 1.60 & 0.525   & 0.530  & 0.676 & 0.30   & 0.972 & 3.382 & 0.29  \\
3D KA  & 1.20 & 0.28    & 0.295  & 0.402 & 0.21   & 0.314 & 1.79  & 0.17  \\
2D KA  & 1.20 & 0.477   & 0.501  & 0.852 & 0.11   & 0.260 &   -   & -     \\
2D MKA & 1.20 & 0.251   & 0.214  & 0.361 & 0.12   & 0.166 &   -   & -     \\
2D R10 & 0.85 & 0.181   & 0.326  & 0.453 & 0.22   & 0.177 &   -   & -     \\
\hline
\end{tabular}
\label{tab:parameters}
\end{center}
\end{table*}

\section{Summary and conclusions}

In this work we have attempted to analyze the breakdown of the
Stokes-Einstein relation in systems of spatial dimensionality 2, 3 and
4, and its relation to the thermodynamics of the system in the form of
the variation of configurational entropy, and various aspects of
dynamical heterogeneity. We summarize here the salient findings of
this work: (i) We find that in systems in all spatial dimensions, the
low temperature relationship between diffusion and structural
relaxation times is well described by a fractional Stokes-Einstein
relationship. (ii) The high temperature behavior is consistent with
the Stokes-Einstein relationship in four and three dimensions, but we
find that in the two dimensional systems we study, the SE relation is
not valid at high temperatures also. In particular, representing the
behavior as a fractional SE relationship, we find the exponent to be
bigger than 1. (iii) We find that the observed breakdown at low
temperatures can be rationalized and understood in terms of the
different activation free energies for diffusion and relaxation times
in three and four dimensions, and in terms of a generalized Adam-Gibbs
relation in two dimensions. (iv) We find that the crossover from high
temperature to low temperature behavior (which constitutes a breakdown
of the SE relation for three and four dimensional systems) occurs
close to the onset temperature of slow dynamics, rather than close to
the mode coupling temperature as previously discussed. (v) We find
that the exponent characterizing the SE breakdown is different for
different systems even for the same spatial dimensions (as seen in two
and three dimensions) calling into question theories which claim
universal exponents for a given spatial dimension. (vi) We find that
the degree of breakdown of the SE relation correlates reasonably well
with measures of dynamical heterogeneity, but with notable exceptions
such as the temperature dependence of the non-Gaussian parameter
$\alpha_{2}$. (vii) We find that the dependence of fragility on
spatial dimensionality is the opposite of the degree of the breakdown
of the SE relation, with fragility becoming larger in higher spatial
dimensions. Although surprising, a strong case for a correlation
between fragility and heterogeneity is not a given, and arguments for
such a correlation in three dimensions need to be examined in light of
our results for their applicability to other dimensions. It would be
interesting to pursue some of the questions raised by our work in
higher spatial dimensions as well, such as recently done in attempting
to identify an upper critical dimension \cite{pap:ucd-Charbonneau}, 
which will hopefully clarify some of the open questions.


\begin{thebibliography}{}
\bibitem{footnote} In order to obtain estimates	of $\beta_{KWW}$, we fit the functions $F_s(k,t)$ to the stretched exponential form $F_s(k,t) = f_{c} \exp(-(t/\tau)^{\beta_{KWW}})$. In order to remove the	influence of the short time decay of	$F_s(k,t)$ on the one hand, and	the noisy long time data on the	other, we fit the $F_s(k,t)$ from the plateau value at the lowest temperature, down to a lower cutoff in the range $0.001 - 0.005$.
\end{thebibliography}
\end{document}